\documentclass[usenatbib]{mnras}

\usepackage{txfonts}
\usepackage[T1]{fontenc}

\DeclareRobustCommand{\VAN}[3]{#2}
\let\VANthebibliography\thebibliography
\def\thebibliography{\DeclareRobustCommand{\VAN}[3]{##3}\VANthebibliography}

\usepackage{graphicx}	% Including figure files

\usepackage[]{xcolor}

\title[On the large scale morphology of Hercules~A: destabilized hot jets?]{On the large scale morphology of Hercules~A: destabilized hot jets?}

\author[]{Manel Perucho,$^{1,2,3}$\thanks{E-mail: manel.perucho@valencia.edu}
Jose L\'opez-Miralles,$^{1,4}$
Nectaria  A. B. Gizani,$^{5}$
Jos\'e Mar\'{\i}a Mart\'{\i}$^{1,2}$,
Bia Boccardi$^{6}$
\\
$^{1}$Departament d'Astronomia i Astrof\'{\i}sica, Universitat de Val\`encia, C/ Dr. Moliner, 50, 46100, Burjassot, Val\`encia, Spain.\\
$^{2}$Observatori Astron\`omic, Universitat de Val\`encia, C/ Catedr\`atic Jos\'e Beltr\'an 2, 46980, Paterna, Val\`encia, Spain. \\
$^{3}$Institut fur Theoretische Physik und Astrophysik, Universitat W\"urzburg, Emil-Fischer-Strasse 31, 97074 W\"urzburg, Germany.\\
$^{4}$Aurora Technology for the European Space Agency, ESAC/ESA, Camino Bajo del Castillo s/n, Urb. Villafranca del Castillo, Villanueva de la Ca\~nada, Madrid, Spain.\\
$^{5}$School of Science and Technology, Hellenic Open University, Building D, Parodos Aristotelous 18, Peribola Patra 26335, Greece.\\
$^{^6}$Max-Planck-Institut fur  Radioastronomie, Auf dem el 69, D-53121 Bonn, Germany.
}

\date{Accepted XXX. Received YYY; in original form ZZZ}

\pubyear{2015}

\begin{document}
\label{firstpage}
\pagerange{\pageref{firstpage}--\pageref{lastpage}}
\maketitle

\begin{abstract}
Extragalactic jets are generated as bipolar outflows at the nuclei of active galaxies. 
%They classified, depending on their morphology, into powerful Fanaroff-Riley II (FRII), and weaker Fanaroff-Riley I (FRI) jets. However, the observations of these objects at large scales show a wide range of possible morphologies, and even radio sources with different structures for both jets.
Depending on their morphology, they are classified as Fanaroff-Riley type I (centre-brightened) and Fanaroff-Riley type II (edge-brightened) radio jets. However, this division is not sharp and observations of these sources at large scales often show intermediate jet morphologies or even hybrid jet morphologies with a FRI type jet on one side and a FRII type jet on the other. A good example of a radio galaxy that is difficult to classify as FRI or FRII is Hercules~A. 
%This source shows bright radio lobes, but without the presence of a strong impact region between the jet and the ambient medium. The
This source shows jets with bright radio lobes (a common feature of FRII type jets) albeit without the hotspots indicative of the violent interaction between the jet and the ambient medium at the impact region, because the jets seem to be disrupted inside the lobes at a distance from the bow shocks surrounding the lobes. In this paper, we explore the jet physics that could trigger this peculiar morphology by means of three-dimensional relativisitic hydrodynamical simulations. Our results show that the large-scale morphological features of Hercules A jets and lobes can be reproduced by the propagation of a relativistically hot plasma outflow that is disrupted by helical instability modes, and generates a hot lobe that expands isotropically against the pressure-decreasing intergalactic medium. We also discuss the implications that this result may have for the host active nucleus in terms of a possible transition from high-excitation to low-excitation galaxy modes.
\end{abstract}


%%%%%%%%%%%%%%%%%%%%%%%%%%%%%%%%%%%%%%%%%%%%%%%%%%

%%%%%%%%%%%%%%%%% BODY OF PAPER %%%%%%%%%%%%%%%%%% 

\section{Introduction}
Extragalactic jets are formed at the surroundings of supermassive black holes in active galactic nuclei (AGN) by means of magneto-hydrodynamical processes that extract rotational energy from the black hole \citep{1977MNRAS.179..433B}. The jets are thought to be accelerated along the collimating region by the toroidal magnetic field \citep[see, e.g.][]{2004ApJ...605..656V,2007MNRAS.380...51K}, and by the conversion of internal into kinetic energy (through the Bernoulli process) at jet expansions \citep[see, e.g.][]{2007MNRAS.382..526P}. Expansion and acceleration cause a drop in the intensity of the jet's magnetic field and thermal pressure, which leads to outflows being kinetically dominated unless dissipative processes \citep[instabilities, shocks, interactions... see, e.g.,][for a review]{2019Galax...7...70P} take over. The large scale morphology of jets is therefore a consequence of the evolution of the flowing plasma along its propagation.

\citet{1974MNRAS.167P..31F} showed the existence of a morphological dichotomy of radio galaxies, with Fanaroff-Riley type I (FRI) being brighter at their centers and becoming dimmer and more symmetric in brightness towards the outer regions, and Fanaroff-Riley type II (FRII) being, in contrast, brighter at their extremes, namely, at the sites of interaction with the interstellar or the warm-hot intergalactic medium (ISM/WHIM). The dichotomy was associated to jet power, with FRII radio galaxies corresponding to the more powerful jets \citep[e.g.,][]{2001A&A...379L...1G}. Nevertheless, this classification, albeit very useful as an initial grouping, has been shown to simplify in excess the complexity of radio source morphologies. Recent observational works have shown that the relation between power and morphology is unclear, mainly at intermediate and low jet powers, with some low power jets showing FRII morphology, or equal power jets showing obviously different morphologies \citep[see, e.g.][]{2021A&A...648A.102V,2019MNRAS.488.2701M,2022MNRAS.511.3250M}. 

The radio galaxy Hercules~A is a good example of this complexity: on the one hand, its relatively powerful jets would be expected 
to develop typical FRII morphologies, but, on the other hand, they exhibit no hotspots at their limbs, and the jets seem to transit from a collimated regime into a disruptive process. The images of this source obtained with different arrays \citep{2002AJ....123.2312S,2003MNRAS.342..399G,2022A&A...658A...5T} show the development of a kink prior to jet disruption. Beyond these disruptive points, at $\sim 100\,{\rm kpc}$ \citep[as derived from VLA and LOFAR observations, e.g.,][–it is noteworthy that this coincides with the cluster density core radius]{1993AJ....106.2218K,2003MNRAS.342..399G,2022A&A...658A...5T}, the jets inflate large, quasi-spherical lobes that resemble those found in FRII radio sources and are surrounded by bow shocks \citep{Nulsen2005}, albeit missing the bright interaction sites, or hotspots. These lobes show, furthermore, notorious arc-shaped filaments. Finally, the western jet/lobe show bow-shock structures that could be triggered with the propagation front of enhanced injection. In summary, the detailed imaging of this source reveals the large level of complexity required to interpret it. 

The jet formation and acceleration paradigm establishes that the jet energy flux is dominated by the magnetic and/or internal energy contributions at injection and it becomes kinetically dominated after becoming superfast magnetosonic \citep[see, e.g.,][]{2012rjag.book...81K}. Entrainment contributes to mass-loading and the correspondent relative increase of kinetic+rest mass energy flux in the flow \citep{2021MNRAS.500.1512A}. Moreover, jet propagation through a decreasing density ambient medium facilitates head acceleration \citep[e.g.,][]{2019MNRAS.482.3718P} and also the gain of flow inertia, which contributes to further jet stability, by increasing the relativistic density relative to the ambient medium \citep[the cocoon expands faster and its density falls faster, too, see e.g.,][]{2012IJMPS...8..241P}. The role of jet inertia in the growth rates of Kelvin-Helmholtz instability (KHI) modes in relativistic flows was discussed in detail in \citet{2005A&A...443..863P}, both for symmetric (pinching) and antisymmetric (helical) modes.

Numerical simulations of extragalactic jets have been traditionally aimed to either reproduce the morphologies of powerful FRII jets \citep[e.g.,][]{1997ApJ...479..151M,1998MNRAS.297.1087K,1999ApJ...523L.125A,2002MNRAS.331..615S,2010MNRAS.402....7M,2016A&A...596A..12M,2016MNRAS.461.2025E,2020MNRAS.499..681M,2021ApJ...920..144S,2011ApJ...743...42P,2014MNRAS.441.1488P,2019MNRAS.482.3718P,2022MNRAS.510.2084P}, or low-power, FRI jets \citep[e.g.][]{2007MNRAS.382..526P,2008A&A...488..795R,2014MNRAS.441.1488P,2016A&A...596A..12M,2019A&A...621A.132M,2022A&A...659A.139M}. Few works have tried to address the complex morphology of Hercules~A-like sources. \citet{2008ApJ...686..843N} simulated the jets in this source as {\it magnetic towers}, obtaining one of the closest approaches to this source found in the literature. 
However, the simulations show jets that are much wider than the observed ones, relative to the lobe size obtained. Furthermore, inherent to the magnetic tower model is the assumption of a Poynting-flux dominated flow, while there are serious doubts about magnetic energy dominating the jet energy flux at these scales 
\citep[see, e.g.][]{1984RvMP...56..255B,2004ApJ...605..656V,2005ApJ...626..733C,2007MNRAS.380...51K}. 
Hence, the question remains about how to 
produce the morphologies seen in the Hercules A jets with non-magnetically dominated flows. \citet{2002ApJ...579..176S} studied the generation of rings in the western lobe of this source via non-relativistic, hydrodynamical simulations. The authors suggested that these rings could be triggered by shocks propagating through the lobes of the disrupted jet and the presence of strong recollimation shocks as the origin of jet disruption and transition to turbulence. However, as we will discuss in this paper, a non-relativistic, axisymmetric approach is probably insufficient to reproduce and interpret correctly the morphology of the lobes in Hercules~A. 

As previously noted, jet disruption by helical instability modes or kinks seems to happen at around the distance at which the cluster density profile starts to fall, after which the quasi-spherical lobes develop. This sphericity is a signature of pressure-driven expansion, which could be favoured by the jets being initially hot \citep{2017MNRAS.471L.120P}. It is very important to remark that when we refer to a \emph{hot} jet, it is not only that its sound speed is large, and its Mach number is thus small, but to the fact that its specific internal energy is relativistic.

In this paper, we try to explain the generation of large scale diffuse lobe morphologies similar to those observed in Hercules~A from a purely relativistic hydrodynamics (RHD) approach, using our code \textsc{Ratpenat} \citep{2010A&A...519A..41P}. Therefore, an attempt to reproduce internal lobe features (rings/arcs) in detail is out of the scope of this work. Interestingly, though, our simulations also reveal a possible physical mechanism responsible for the formation of the observed arcs in Hercules~A. From our results, we suggest that these features could be the consequence of the development of (helical) instability-triggered shocks. Future work will revisit this scenario with relativistic magneto-hydrodynamics (RMHD) simulations using the code \textsc{L\'ostrego} \citep{2022A&A...661A.117L,2023CoPhC.28408630L}. As a warning note, we point out that it is not the aim of this paper to reproduce the exact morphology of the source, which would require not only a detailed knowledge of the jet injection conditions, including the jet power, rest-mass density and velocity, specific internal energy, and the structure of the magnetic field, but also of the environment prior to jet injection. It would be therefore reckless to approach the problem with this aim. Alternatively, we adopt a qualitative approach, which allows us to discard some scenarios and to put the focus on particular jet/ambient medium configurations to explain both jet disruption and lobe inflation.

We have run numerical simulations of jets with a set of parameters that are based on two different hypothesis about the origin of the observed morphology. In particular, the two possibilities considered are: 1) a kinetically dominated jet, with a periodic variation in injection power, and 2) an internal-energy dominated jet, prone to the development of instabilities and disruption. In the first case, the lobes would be inflated by the shocked jet plasma injected prior to the drop in jet power, whereas in the second this task would be done by the continuously injected, high-pressure plasma going through the disruption point. We show that a jet with a large internal energy 
(with a possibly contribution of magnetic energy) can produce a qualitatively similar morphology to that in Hercules~A, avoiding the assumption of Poynting-flux dominated, force-free jets at kpc scales, 
not favoured by current models of jet acceleration \citep{2012rjag.book...81K} and evolution \citep{2019Galax...7...70P}. Owing to the large computational cost of three-dimensional numerical simulations, we are forced to run only a few simulations, focusing on fitted density/pressure profile of the cluster from observational results \citep{2004MNRAS.350..865G} and basic estimates on the jet properties \citep{2003MNRAS.342..399G}. A broad sweep in parameter space is prohibitive in 3D simulations like the ones we present here, which require between $5\times10^5$ and $10^6$ hours of calculation in supercomputing resources.

The structure of the paper is the following: Section~\ref{sec:num} is devoted to the description of the simulation setup; in Section~\ref{sec:results} we describe our results, and in sections~\ref{sec:disc} and \ref{sec:conc} we discuss those results and present our conclusions, respectively.

\section{Numerical simulations}\label{sec:num}

We have used the code \textsc{ratpenat} running in the supercomputing resources at the University of Val\`encia. This code solves the 
%equations of relativistic hydrodynamics 
RHD equations in conservation form, using high-resolution-shock-capturing methods \citep[see][]{2010A&A...519A..41P}. The code is a hybrid parallel code  -- MPI + OpenMP -- that is optimized for the use of supercomputers with a shared/distributed memory structure. The simulations have been run in 256 to 1024 cores at Tirant (\emph{Servei d'Inform\`atica de la Universitat de Val\`encia}). We have used LLNL VisIt \citep{Childs} to generate the figures for this work.

We also introduced a source term in the energy equation to account for thermal cooling in a simplified way \citep{2017A&A...606A..40P, 2022MNRAS.510.2084P}. The cooling term $\Lambda$ (cgs units; considering hydrogen ions only) is taken as in the approximation given in \cite{1998MNRAS.298.1021M}:

\begin{equation}
 \Lambda\,=\, n_e\,n_H \times
 \left\{ 
 \begin{array}{lr}
 7\times10^{-27}T,& 10^4\leq T \leq 10^5  \\  \nonumber
 7\times10^{-19}T^{-0.6},& 10^5\leq T \leq 4\times10^7 \\
 3\times10^{-27}T^{0.5},& T \geq 4\times10^7    \nonumber
 \end{array} \right\} \,
\end{equation}

\noindent
where $T$ is the gas temperature, $n_e$ and $n_H$ are the electron and hydrogen ion number densities. Whereas $n_e = \rho_e/m_e$ with $\rho_e$ the electron mass density and $m_e$ the electron mass, for purely ionized hydrogen, $n_H = \rho_p/m_p$ with $\rho_p$ the proton mass density and $m_p$ the proton mass. Electron and proton mass densities are computed with the Synge equation of state, implemented as described in Appendix~A of \citet{2007MNRAS.382..526P}, which allows us to describe our system as a mixture of relativistic (electron, proton and positron) ideal gases. We therefore implicitly assume thermodynamical equilibrium at each cell.

\subsection{The ambient medium}
%\subsection{Set up}\label{sec:set up} % used for referring to this section from elsewhere

The numerical box used in the simulations defines a cube with sides of 256 or 512~kpc, with the jet injected from the center of one of its faces as a boundary condition. The jet injection point is taken to be at $\sim\,10\,{\rm kpc}$ from the active nucleus, as fixed by the distribution of the ambient medium (see below). The jet radius at injection is taken to be $R_j\,=\,1\,{\rm kpc}$. The ambient medium is given by a density profile that corresponds to that estimated for Hercules~A \citep[see, e.g.,][]{2004MNRAS.350..865G}. We model it as a hot medium dominated by ionised hydrogen (electron-proton gas) with the following profile

\begin{eqnarray}\label{next}
  n_{\rm ext} = n_{\rm c} \left(1 +
\left(\frac{r}{r_{\rm c}}\right)^2\right)^{-3\beta_{\rm atm,c}/2},
\end{eqnarray}
where $r$ is the radial spherical coordinate. Here, we have used $n_{\rm c} = 0.01$~cm$^{-3}$, $r_{\rm c} = 120$~kpc, and $\beta_{\rm atm,c} = 2.22$. The temperature is set constant to $4.9\times10^7$~K. The pressure is derived from
\begin{equation}\label{pext}
  p_{\rm ext} = \frac{k_{\rm B} T_{\rm ext}}{\mu X} n_{\rm ext},
\end{equation}
where $\mu$ is the mass per particle in atomic mass units ($\mu=0.5$ here), $X$ is the abundance of hydrogen per mass unit, which is set to 1, and $k_{\rm B}$ is the Boltzmann's constant. 

Note that the chosen profiles for the density and temperature define a dilute and hot ambient medium. Whereas this can be the result of the passage of the jets (which sweep and heat the original ambient medium as they propagate), the lack of hotspots and strong bow shocks around the radio source at the simulated scales, along with the long lifetimes estimated for the source would suggest that the ambient medium has got enough time to recover its original ditribution ($t_{\rm jet} > t_{\rm G}$, the dynamical time of the gravitational pull from the host galaxy). Nevertheless, it is impossible to know the initial distribution of the WHIM prior to the injection of the radio galaxy.

\begin{table*}
	\centering
	\caption{Set up parameters in the different simulations. The columns indicate, in order, the model identification (K for kinetically dominated jets and I for internal energy dominated jets), box size, jet total kinetic power, jet rest mass-density, flow velocity and enthalpy at injection, gradual switch off time, modulation periodicity and simulated time.}
	\label{tab:setup}
	\begin{tabular}{cccccccccc} 
    	& Cells & Size (kpc$^3$) & $L_{\rm j}$ (erg/s) &  $\rho_{\rm j}$ (g/cm$^{3}$) & $v_{\rm j}$ ($c$) & $h_{\rm j}$ ($c^2$) & $t_{\rm off}$ (Myr) & $T_{\rm mod}$ (Myr) & $t_{\rm end}$ (Myr)\\ 
		\hline
%	 K 	&  &  &  &  & &  & & & \\ 
	K - Case 1 & $512^{3}$ & $256^3$ & $2\times 10^{46}$ & $1.67\times 10^{-30}$ & 0.92 & 2.6 & - & - & 41\\ 

	K - Case 2 &$512^{3}$ & $256^3$ & $2\times 10^{46}$ & $1.67\times 10^{-30}$ & 0.92 & 2.6 & 19 & - & 43 \\  
		
    K - Case 3 &$512^{3}$ & $256^3$ & $4\times 10^{45}\,-\,2\times 10^{46}$ & $1.67\times 10^{-30}$ & $0.83\,-\,0.92$ & 2.6 & 19 & 10 & 30\\
        \hline
 I  & $1024^{3}$ & $512^3$ & $4\times 10^{45}$ & $1.67\times 10^{-33}$ & 0.92 & 537.3 & - & - &  152\\
          \hline
	\end{tabular}
\end{table*}

\subsection{Kinetically dominated jets}
The jet power, in the absence of a dynamically relevant magnetic field, is given by $L_j = \rho(h \gamma - 1) \gamma c^2  v  A_j$, with $\rho$ the rest mass density, $h$ the specific enthalpy, $v$ the jet flow speed and $\gamma$ the corresponding Lorentz factor, $A_j$ the jet area at injection, and the subtraction in the bracket removes the contribution of the rest-mass energy of the particles. The specific enthalpy is related with the specific internal energy $\varepsilon$ through  $h = 1 + \Gamma \varepsilon/c^2$ where $\Gamma$ is the adiabatic index provided by the equation of state.

As pointed out in \citet{2017MNRAS.471L.120P}, the amount of internal energy transported by the jet affects the lobe pressure, and it is precisely when $\varepsilon \gg c^2$ that the jet energy flux is dominated by internal energy. On the contrary, when $\varepsilon \sim c^2$ or smaller, the jet is dominated by kinetic energy.

The injection power of the kinetically dominated jet is $L_j\,=\,2\times10^{46}\,{\rm erg/s}$, within the estimated range of jet power for the source \citep{2002ApJ...579..176S}. The jet was injected into the ambient medium grid (a numerical box with 512$^3$ cells, covering a physical region of (256~kpc)$^3$) as kinetically dominated, with velocity $v\,=\,0.92\,c$ and rest mass density $10^{-4}$ times that of the ambient medium at injection, i.e., $\rho_j\,=\,1.67\times10^{-30}\,{\rm g/cm^3}$. The leptonic contribution to the total density is taken to be of the 80\%. By doing this we account for possible entrainment of protons along the initial kiloparsecs of evolution, as expected from, at least, jet/star interactions \citep[see, e.g.,][]{2021MNRAS.500.1512A}. The jet specific internal energy is $\varepsilon\,=\,1.15\,c^2$, which makes the jet also thermodynamically relativistic, but still dominated by kinetic energy. The adiabatic exponent of the flow at injection is $\Gamma\,=\,1.39$. The classical Mach number of the jet at injection is thus 1.88, whereas the relativistic Mach number is 4.8.

In this first set-up, we fixed constant injection up to $t\,\simeq\,19\,{\rm Myr}$, which is the time at which the bow shock has crossed the density core, $z_{\rm BS}\,\simeq 125\,{\rm kpc}$. After that point, three different simulations were run as a continuation of the initial one: 1) uninterrupted, constant injection 2) a gradual switch off of injection after $t_c$, and 3) periodical modulation of the injection velocity between $0.83\,c$ (dropping jet power to $L_j\,=\,4\times10^{45}\,{\rm erg/s}$) and $0.92\,c$, with a periodicity of $10\,{\rm Myr}$. This period was chosen following the hypothesis that the observed arcs in Hercules~A are triggered by patterns, which could propagate at smaller velocities than the jet flow and turn into shocks when moving through the slower lobe plasma \citep[see, e.g.,][for a discussion on instability generated shocks]{2005A&A...443..863P}. In addition, we chose velocity values for the modulated injection that resulted in jet powers between the limits estimated by \citet{2002ApJ...579..176S} ($2\times 10^{45} - 2\times 10^{46}$~erg/s). We want to stress that the introduction of such modulation was mainly aimed to increase the internal energy of the flow and not so much to produce the observed arcs. 

%In all simulations we introduced a helical injection pattern in velocity with different periodicities \citep[see Equation 13 in][]{2019MNRAS.482.3718P}, sweeping from $\sim 4\times 10^{4}$ to $\sim 2\times 10^6\,{\rm yr}$ in these simulations. 
In all these simulations we introduced a helical perturbation pattern in velocity \citep[see Equation 13 in][]{2019MNRAS.482.3718P} with different periodicities ranging from $\sim 4\times 10^{4}$ to $\sim 2\times 10^6\,{\rm yr}$. The total amplitude given to these helical motions is $2.5\times10^{-4}$ of the injection velocity. These helical patterns allow us to break symmetries and trigger KHI helical modes if the flow is prone to their development \citep[see, e.g.,][]{2006A&A...456..493P}.

\subsection{Internal energy dominated jet}
In the case of the internal energy dominated outflow, jet injection velocity is $v_j\,=\,0.92\,c$ -with the helical injection pattern mentioned above-, the gas rest-mass density ratio with the ambient medium is $10^{-7}$, i.e., $\rho_j\,=\,1.67\times10^{-33}\,{\rm g/cm^3}$, with the same composition as the kinetically dominated jet (80\% of the mass as electrons and positrons, and 20\% as protons); the jet specific internal energy is $\varepsilon\,=\,404\,c^2$, and the adiabatic exponent is $\Gamma\,=\,1.33$. The total jet power is, in this case, $L_j\,=\,4\times 10^{45}\,{\rm erg/s}$, within the estimated range of jet power for the source \citep{2002ApJ...579..176S}. This simulation was run in a numerical box with 1024$^3$ cells, with physical dimensions of (512~kpc)$^3$. Table~\ref{tab:setup} summarizes the jet parameters for the four simulations.

\section{Results}\label{sec:results}

\begin{figure*}%
    
	\includegraphics[width=0.33\linewidth]{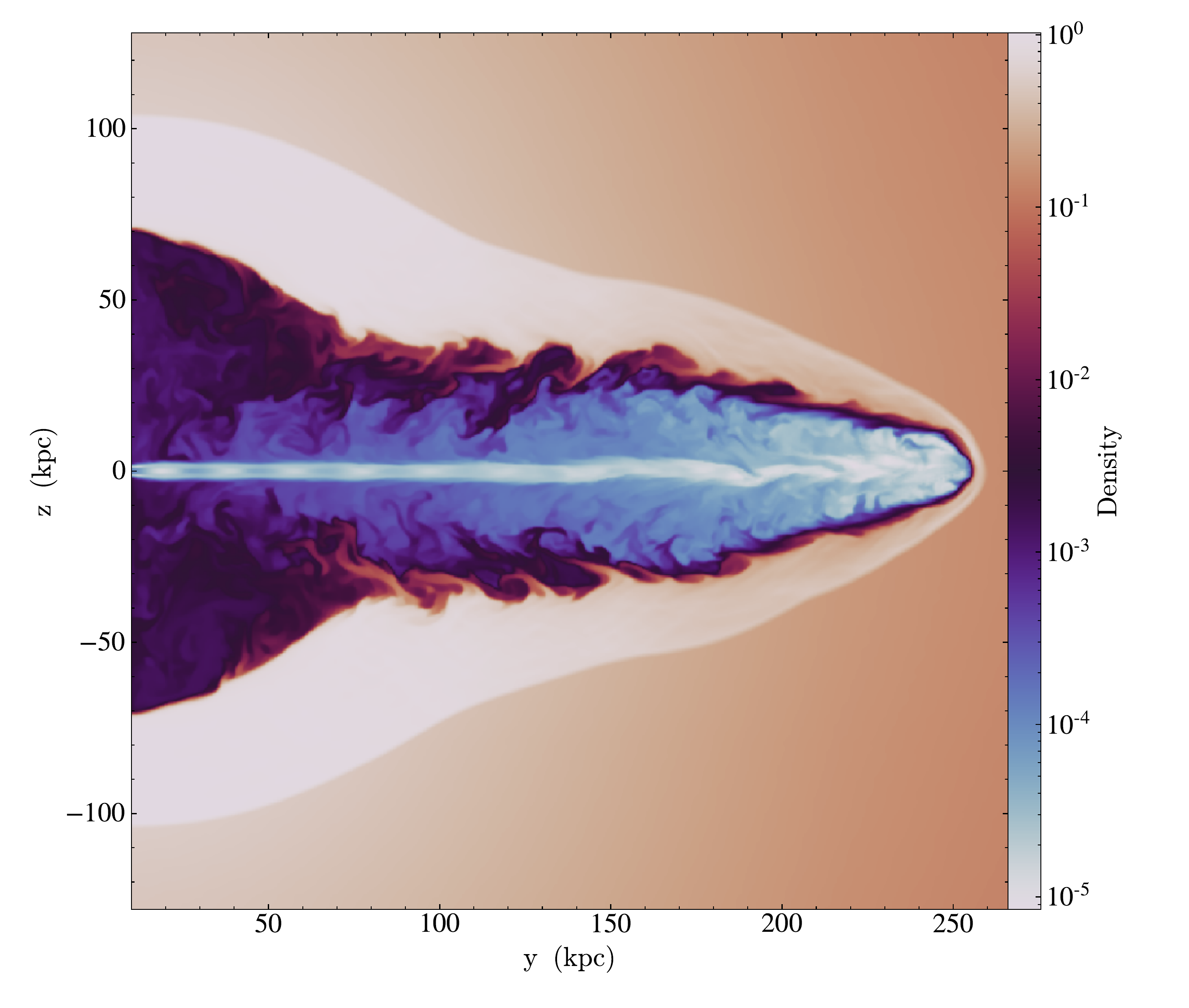}~    
    \includegraphics[width=0.33\linewidth]{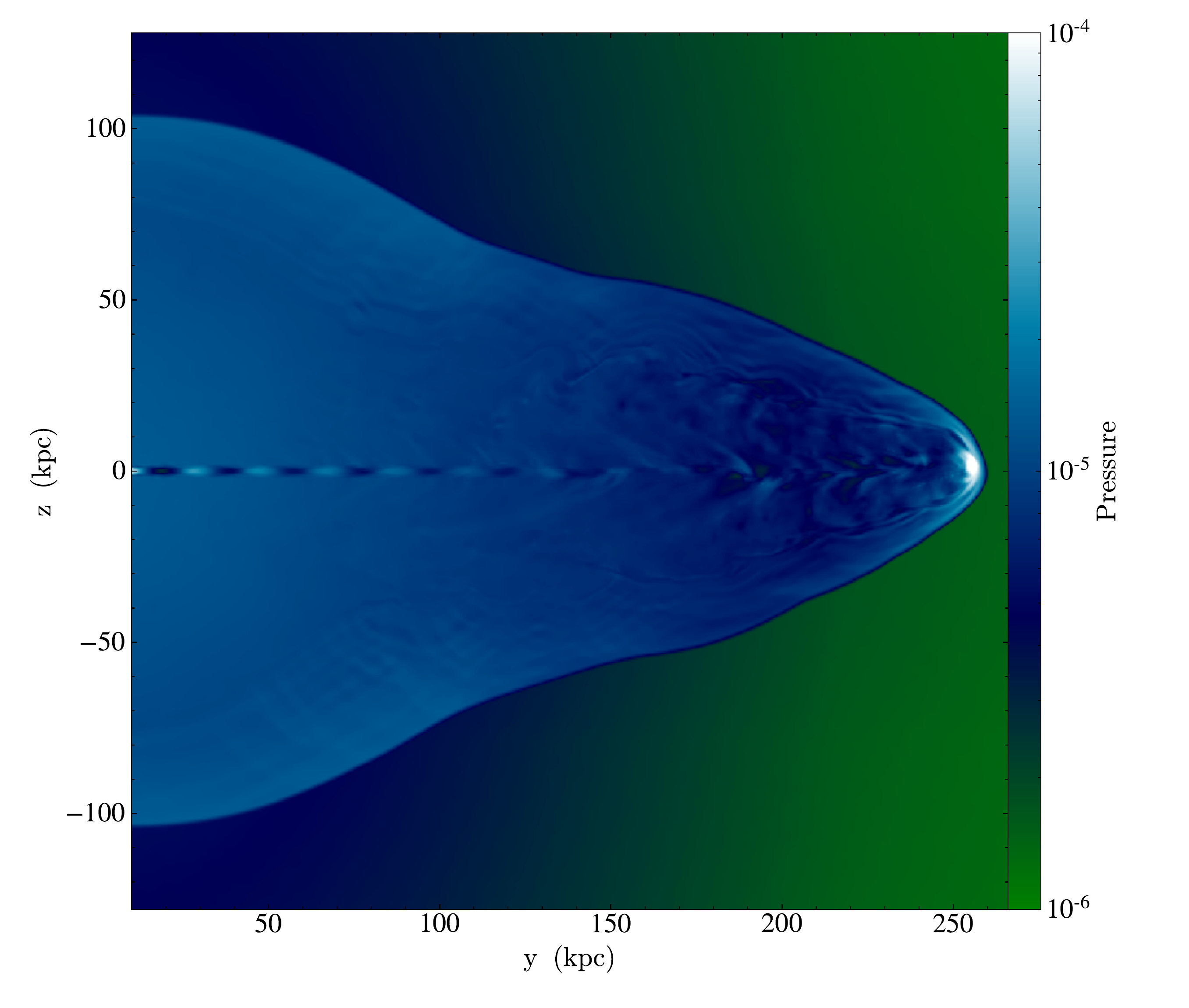}~    
    \includegraphics[width=0.33\linewidth]{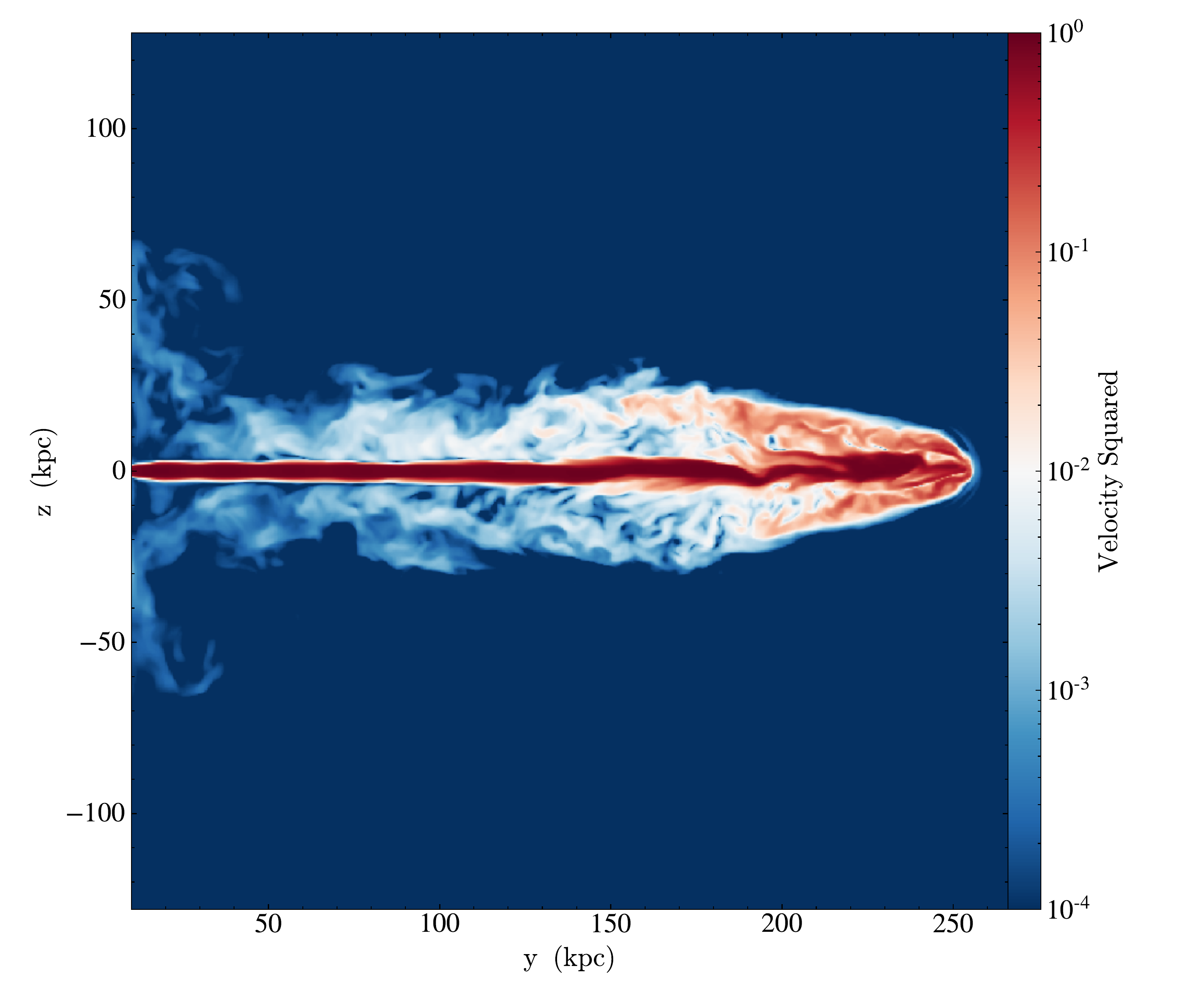}
        
	\includegraphics[width=0.33\linewidth]{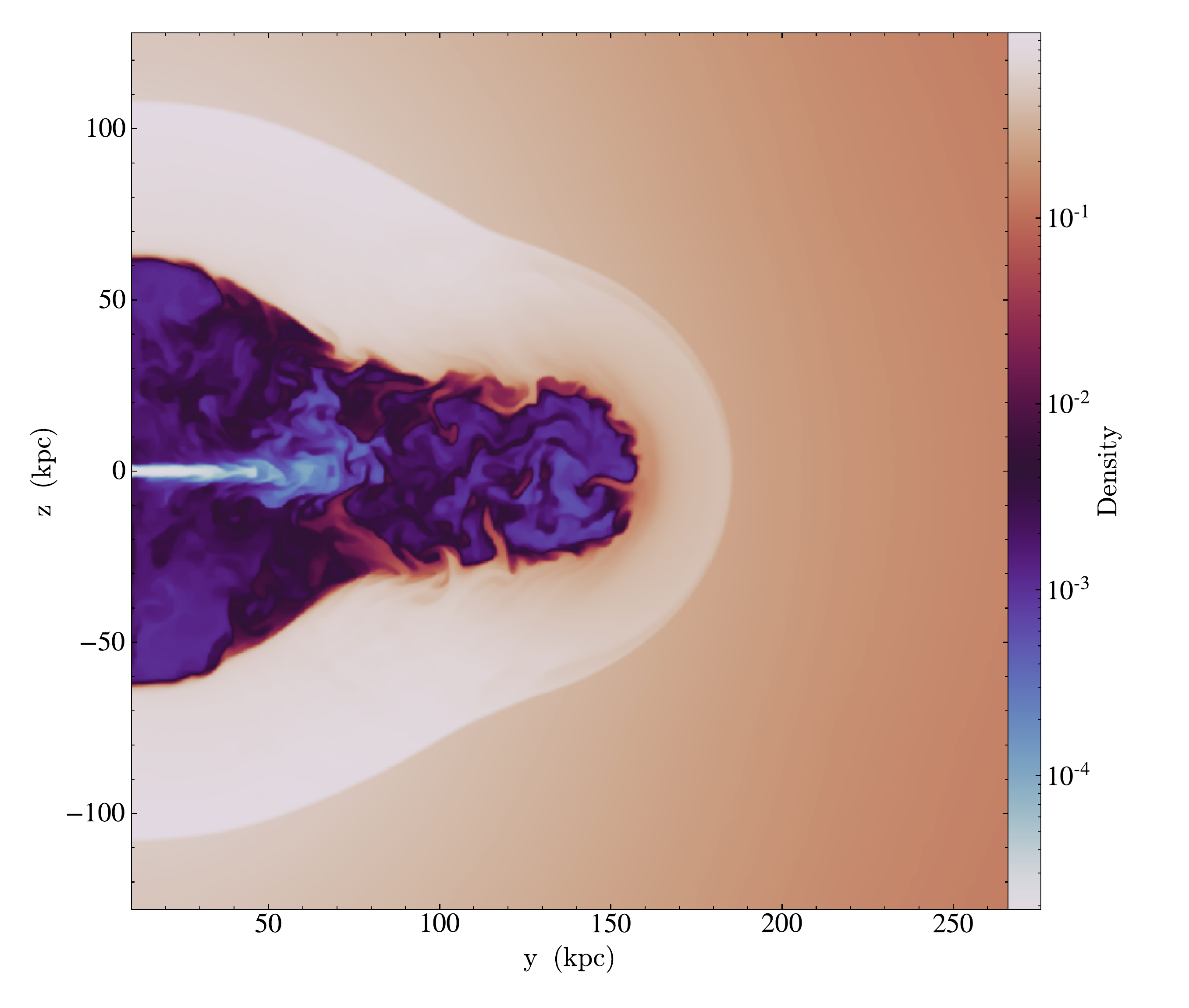}~    
    \includegraphics[width=0.33\linewidth]{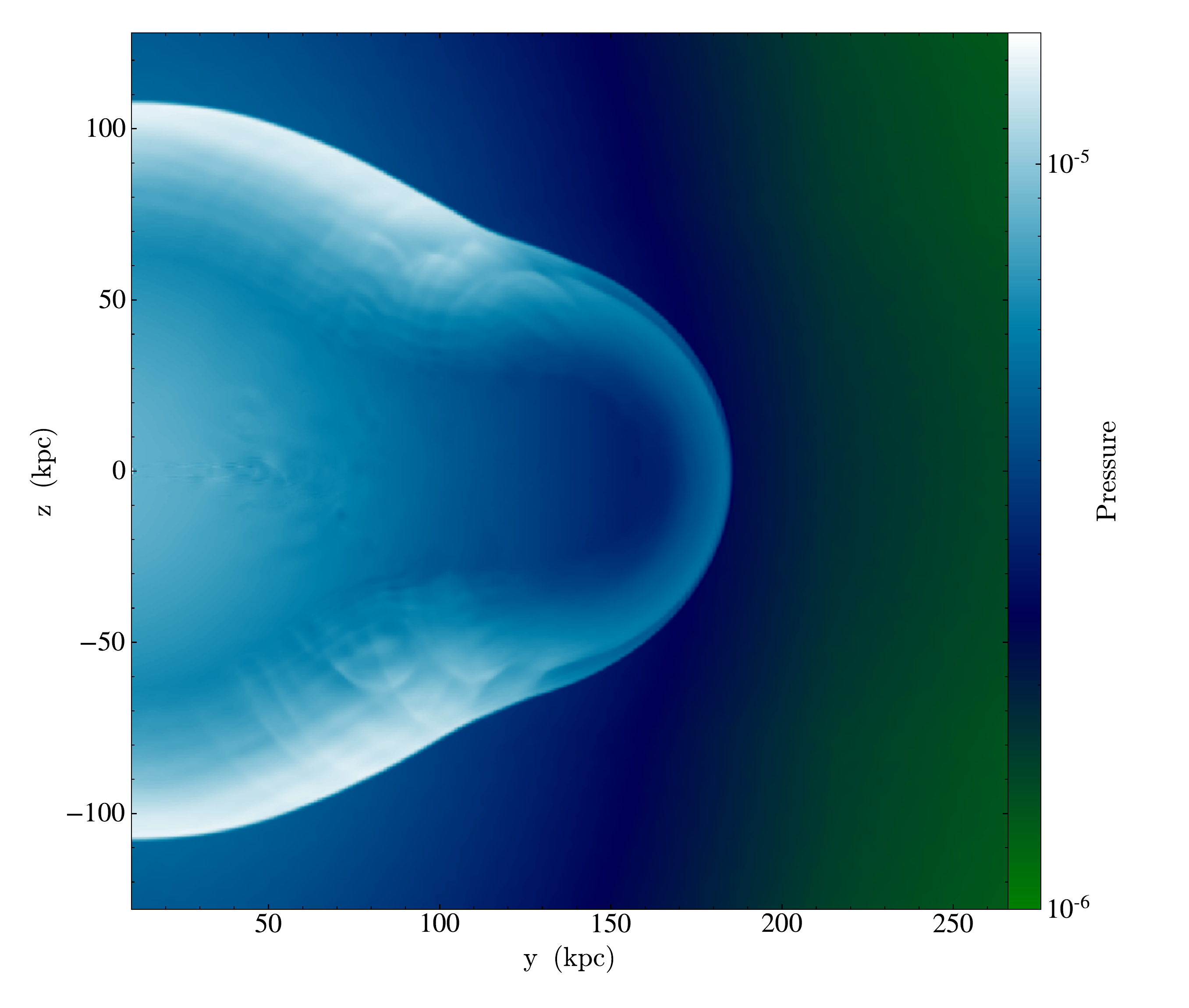}~    
    \includegraphics[width=0.33\linewidth]{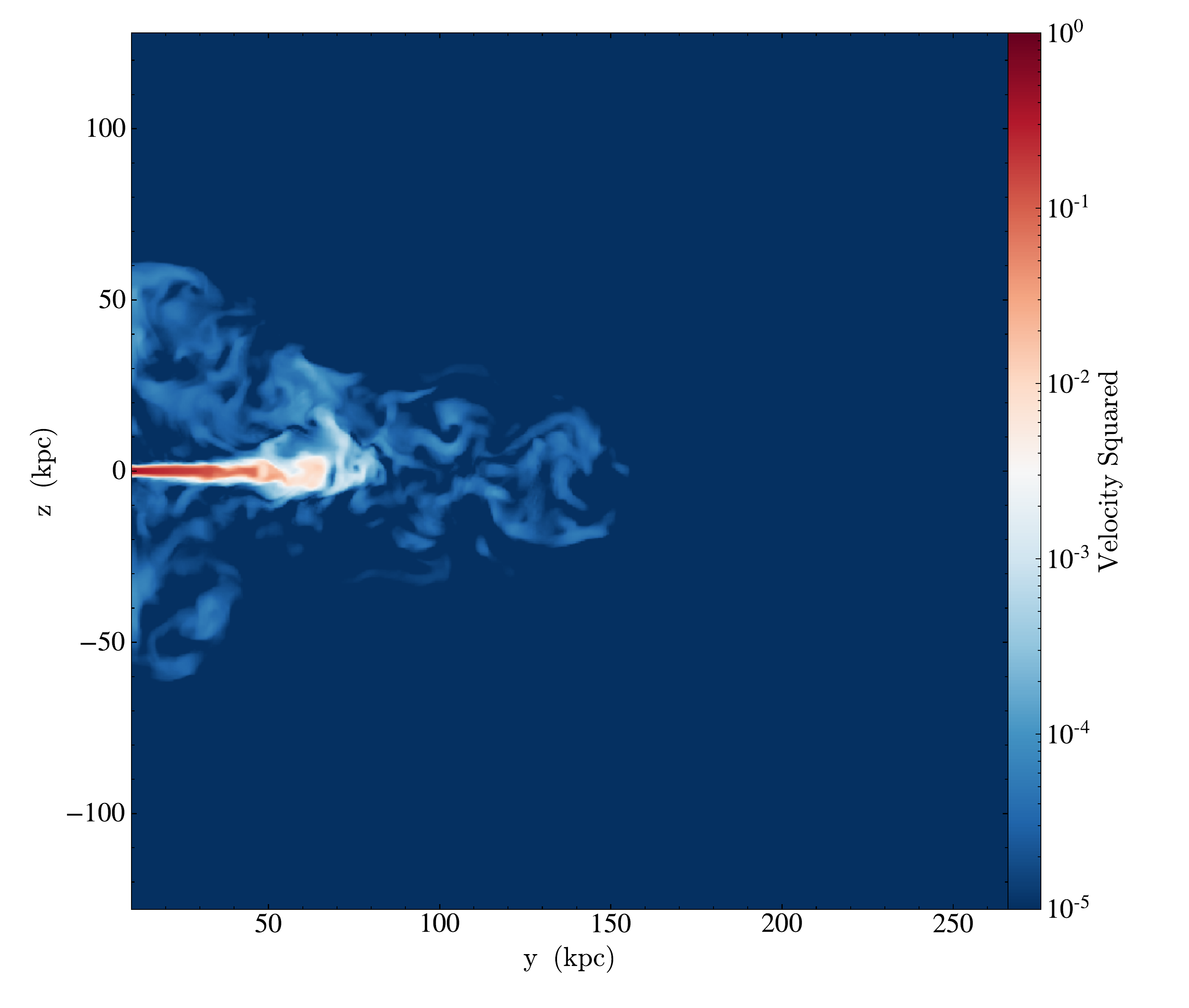}

    \includegraphics[width=0.33\linewidth]{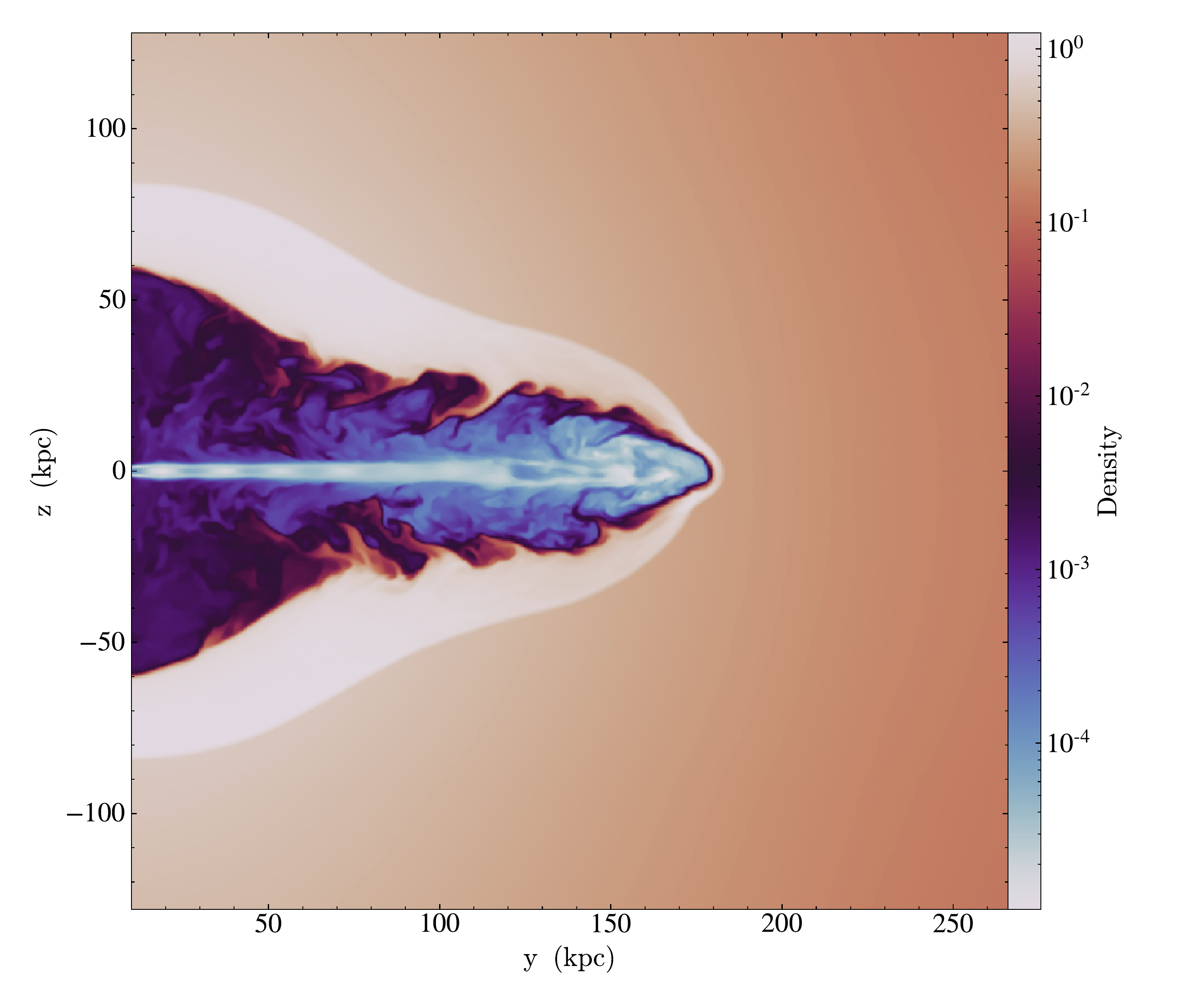}~    
    \includegraphics[width=0.33\linewidth]{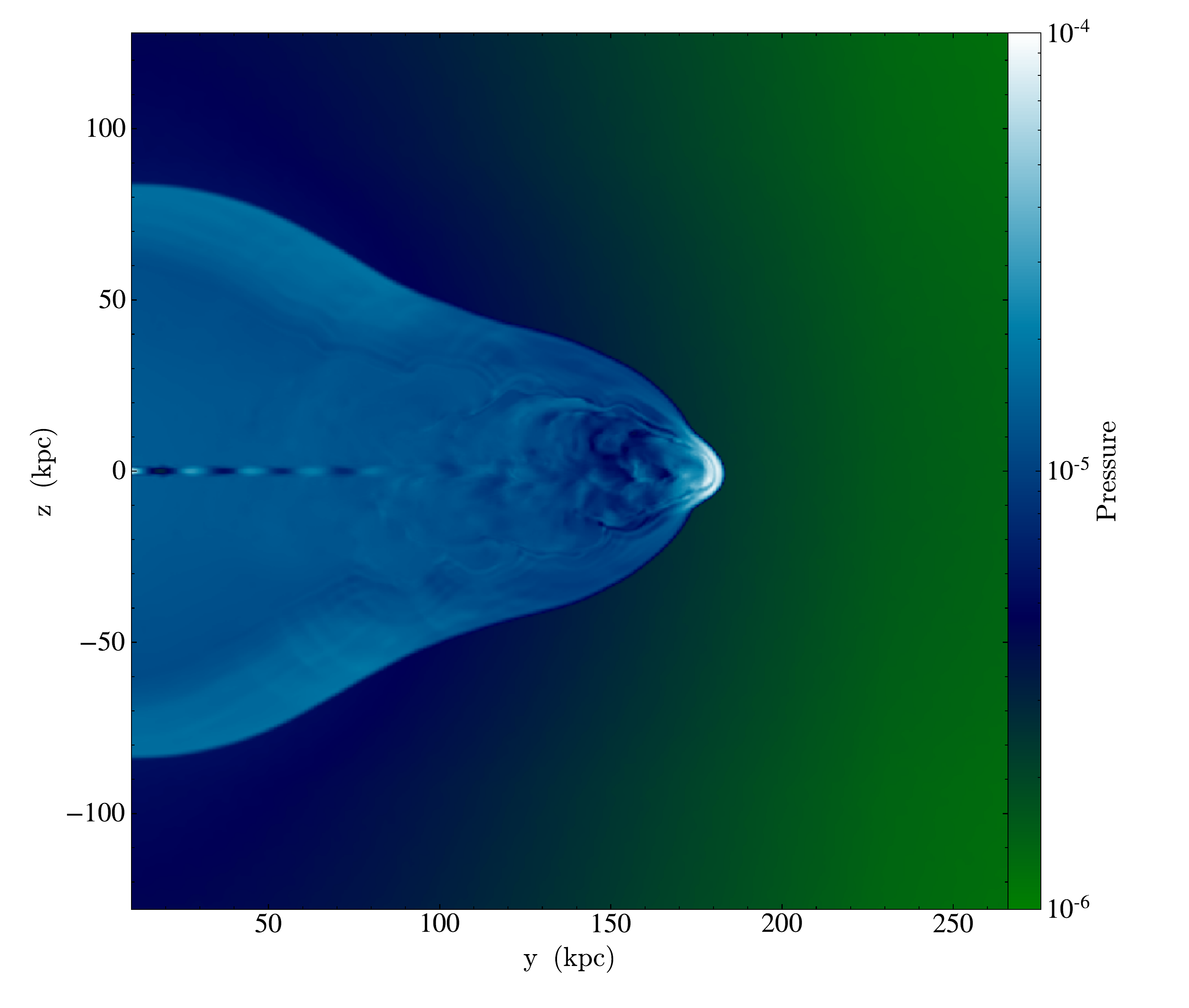}~    
    \includegraphics[width=0.33\linewidth]{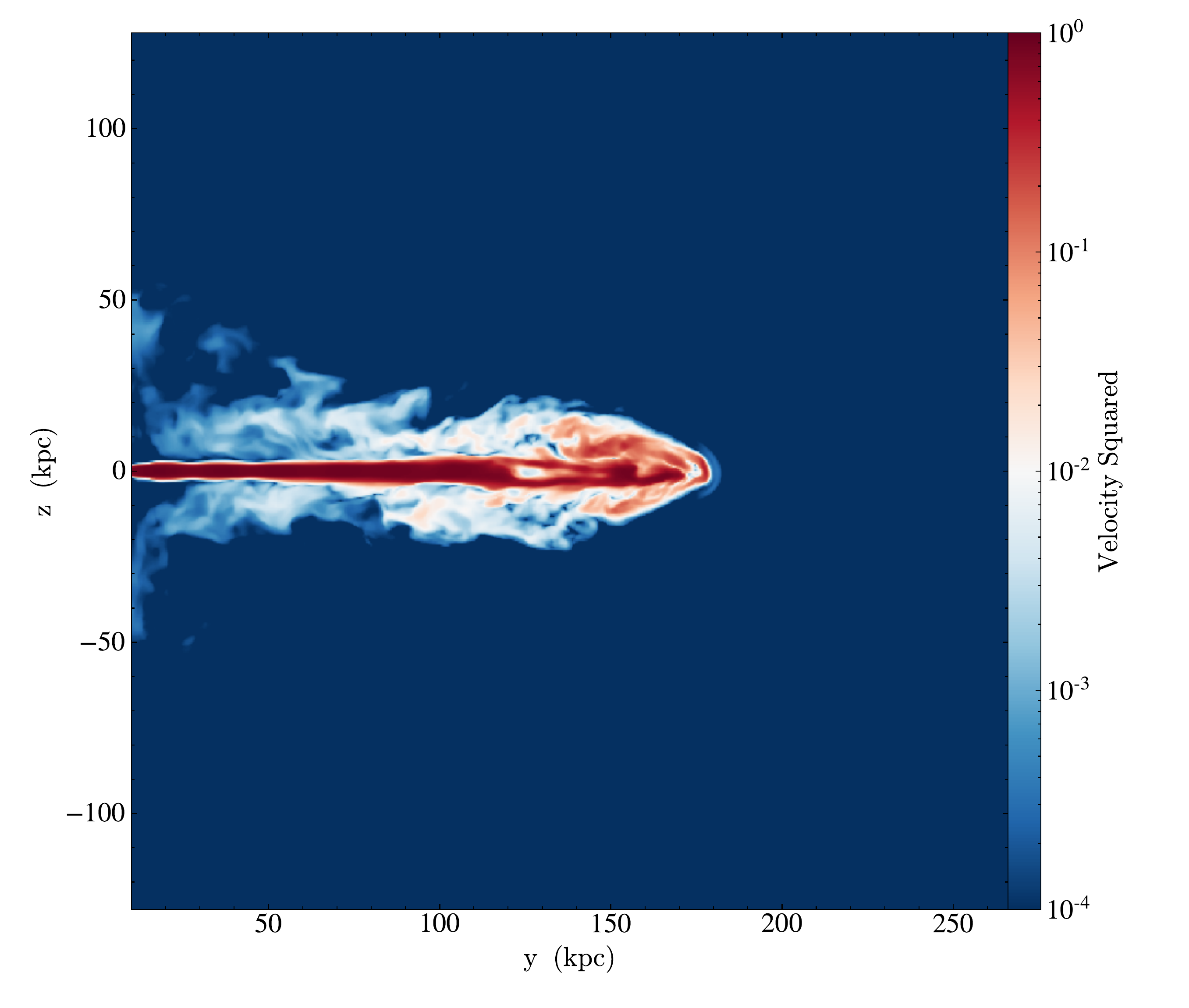}

    \caption{Rest-mass density (left panels), pressure (central panels) and velocity module squared (right panels) cuts for the three kinetically dominated jets at the last snapshots ($t\simeq 41$, $43$, and $30\,{\rm Myr}$, respectively). The top row shows case 1, 
    constant injection, the central row shows case 2, with a gradual switch off, and the bottom panel shows case 3, with modulated injection. 
    %The units are code units ($\rho_a$ for density, $\rho_a\,c^2$ for pressure, and $c$ for velocity).
    Units are: $\rho_a$ for density, $\rho_a\,c^2$ for pressure, and $c$ for velocity.}
  \label{fig:kinetic}
\end{figure*}

\subsection{Kinetically dominated jets}
\begin{figure}%
	\fbox{\includegraphics[width=0.95\columnwidth]{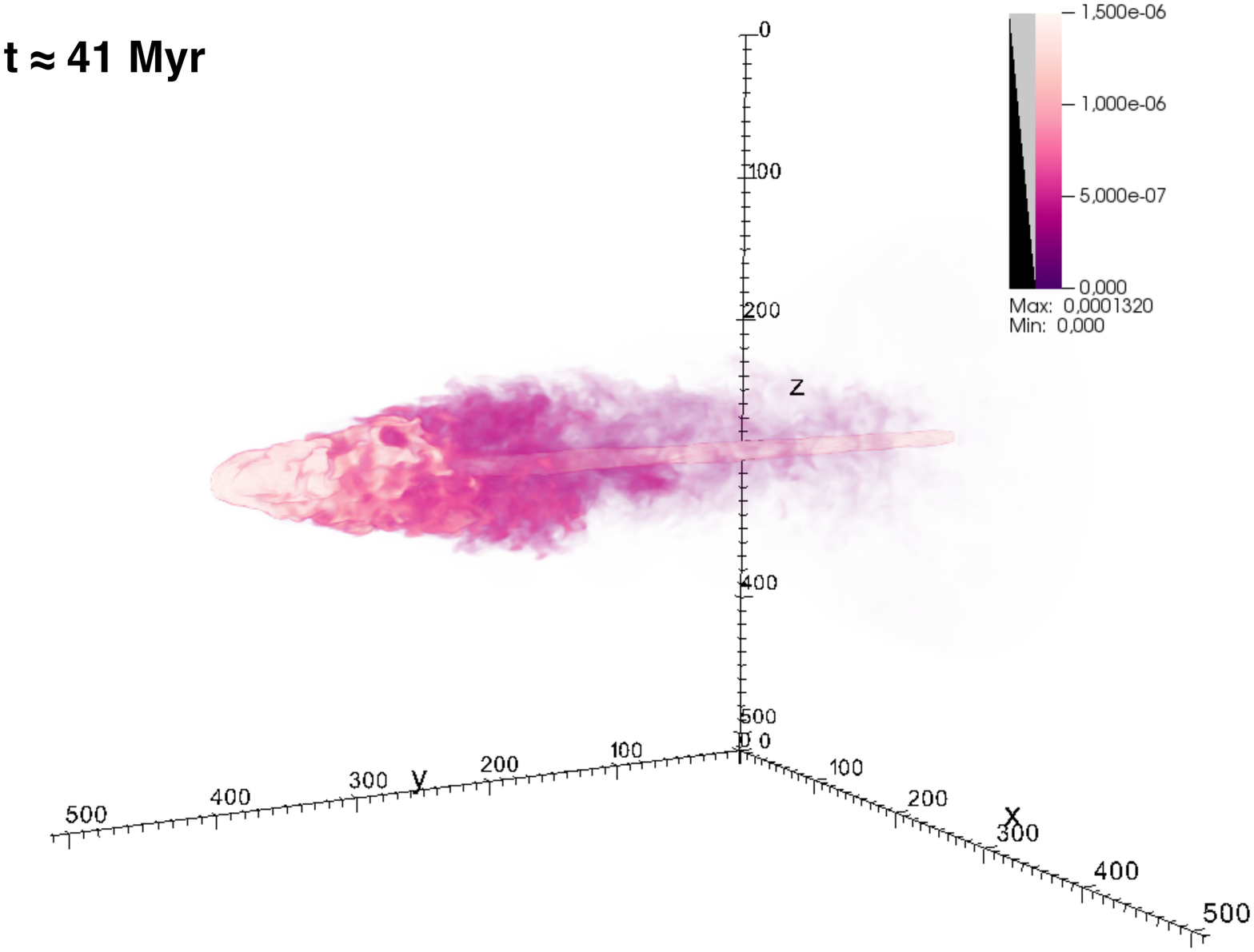}}
 
    \vspace{0.05cm}
    
    \fbox{\includegraphics[width=0.95\columnwidth]{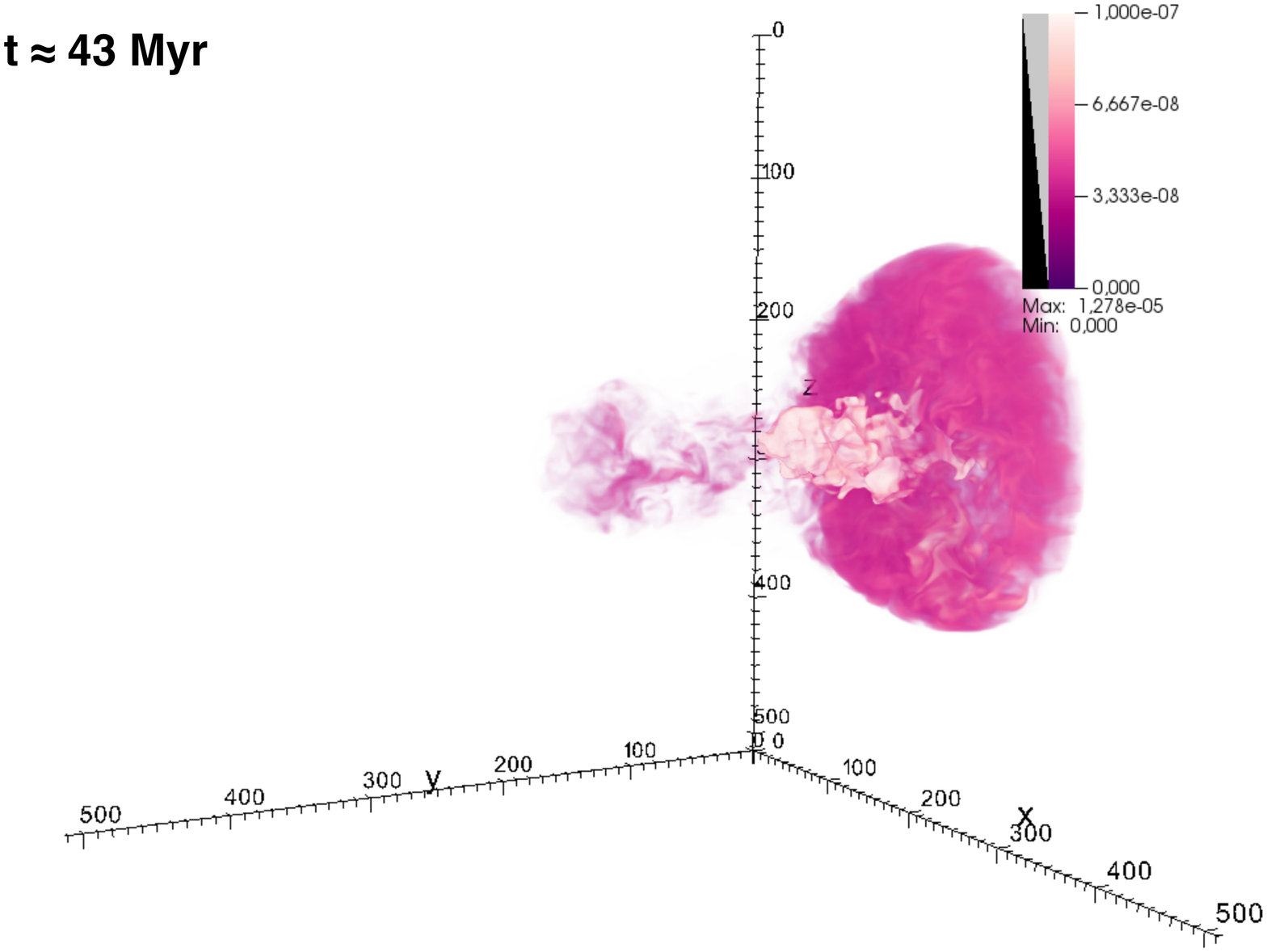}}
    
    \vspace{0.05cm}
    
    \fbox{\includegraphics[width=0.95\columnwidth]{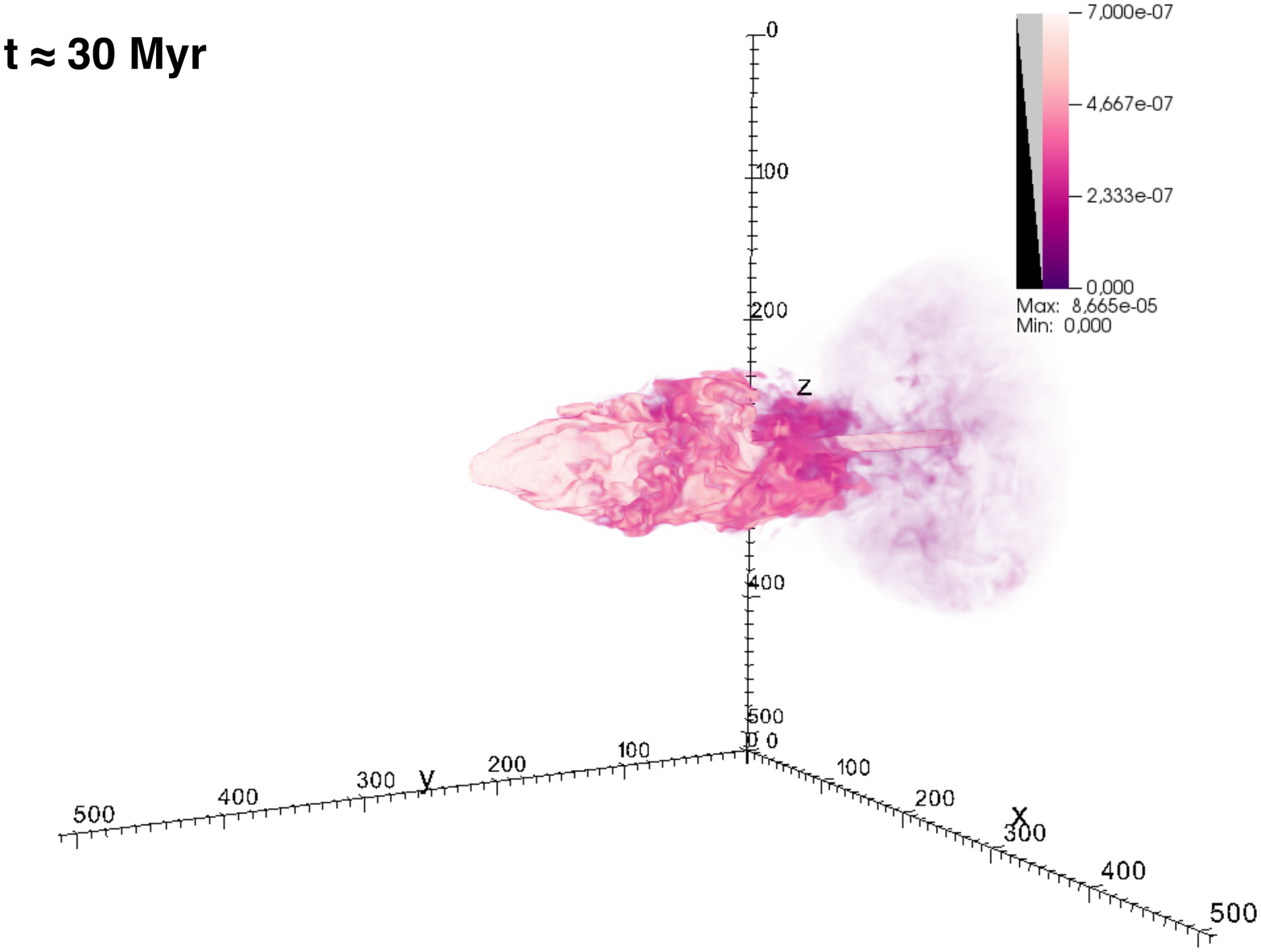}}
    \caption{Pressure (weighted with tracer) renderings of the numerical box for the three kinetically dominated jets, cases 1 to 3 from top to bottom, see text. We apply limits to the range of values in the box, which permits to visualize the regions of interest -jets and lobes. The axes indicate cell numbers.}
  \label{fig:kinetic2}
\end{figure}

Figure~\ref{fig:kinetic} shows cuts of density, pressure and velocity module squared, for the aforementioned simulations 1, 2 and 3, at the last snapshots. 
%Continuous 
Constant injection -case 1, top panels- produces a very different lobe morphology to that observed in Hercules~A, and very similar to the typical structures obtained in long-term numerical simulations of powerful, kinetically dominated jets \citep[e.g.,][]{1997ApJ...479..151M,1998MNRAS.297.1087K,1999ApJ...523L.125A,2002MNRAS.331..615S,2010MNRAS.402....7M,2016A&A...596A..12M,2016MNRAS.461.2025E,2020MNRAS.499..681M,2021ApJ...920..144S,2011ApJ...743...42P,2014MNRAS.441.1488P,2019MNRAS.482.3718P,2022MNRAS.510.2084P}. Although the jet shows periodical expansions and recollimations, visible in the pressure map, the jet disruption is clearly caused by helical distortions of the jet, as seen in the density, left panel. The gradual switch off –case 2, central panels– produces a detached lobe, but this lobe does not develop a spherical shape ($\sim 100$~kpc and $\sim 50$~kpc in the axial and transversal directions, respectively), unlike the observed lobes in Hercules~A. Furthermore, the jet is not observed to enter the lobes as it happens in the VLA images of the source.

The third simulation -bottom panels in Fig.~\ref{fig:kinetic}- involves a periodical modulation of injection power without completely switching it off. At the last snapshot, the jet power is close to its minimum, with $v_j\,\simeq\,0.83\,c$. This set up develops turbulent, inhomogeneous lobes. The resulting structure shows a pinching structure in the jet and strong irregularities caused by the interaction between the different enhanced injection epochs. However, no strong internal shocks are observed, which could be explained by the smoothness of the change in jet power. \citet{2022A&A...658A...5T} have concluded that the periodicity in the jet activity could probably be more than an order of magnitude shorter than the one we have used. Such a short period could result in a more continuous jet-lobe structure, and generate the bow shock structures observed in the western jet, if the changes are significant enough along the activity cycle. Perhaps, a more abrupt change would facilitate shock formation and the generation of arcs, but the morphology obtained from this simulation seems to be far away from the observed radio source. New simulations should be run to test this scenario properly for the case of hot jets, as we show in the next sections. This is however, out of the scope of this paper, which is focused on the generation of the peculiar large scale, jet-lobe morphology of Hercules~A.

Figure~\ref{fig:kinetic2} shows renderings of pressure weighted with tracer for the three cases, for the same snapshots shown in Fig.~\ref{fig:kinetic}. These images highlight the high energy density regions, i.e., those associated to the jet or shocked jet gas, expected to have higher synchrotron emissivity. Altogether, these simulations show that kinetically dominated jets tend to form elongated lobes, even if jet injection is modulated or interrupted \citep[as shown in previous work,][]{2014MNRAS.445.1462P}, or fail to reproduce the observed jet-lobe connection (case 2). 

\begin{figure*}%
    
	\includegraphics[width=0.33\linewidth]{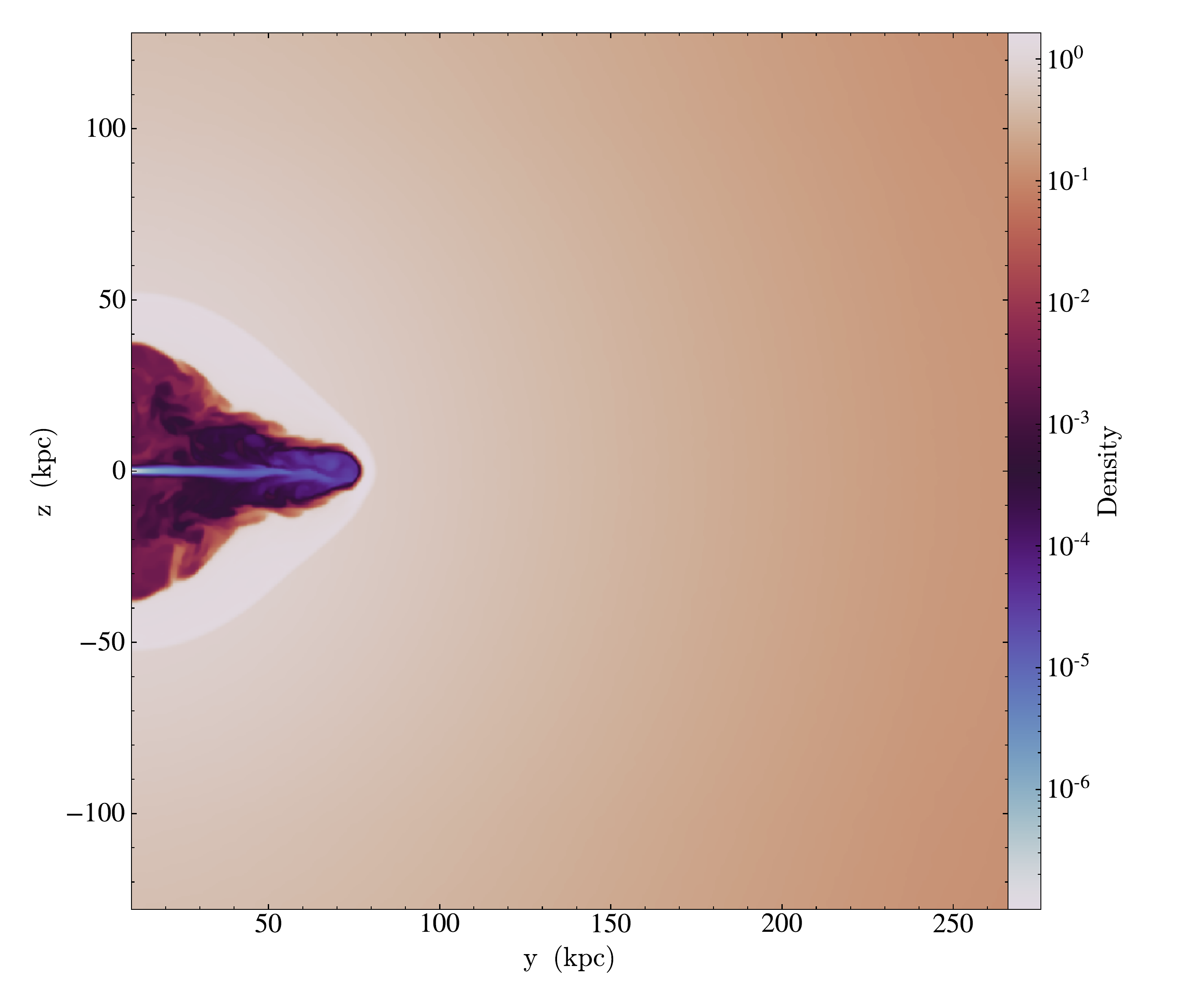}~    
    \includegraphics[width=0.33\linewidth]{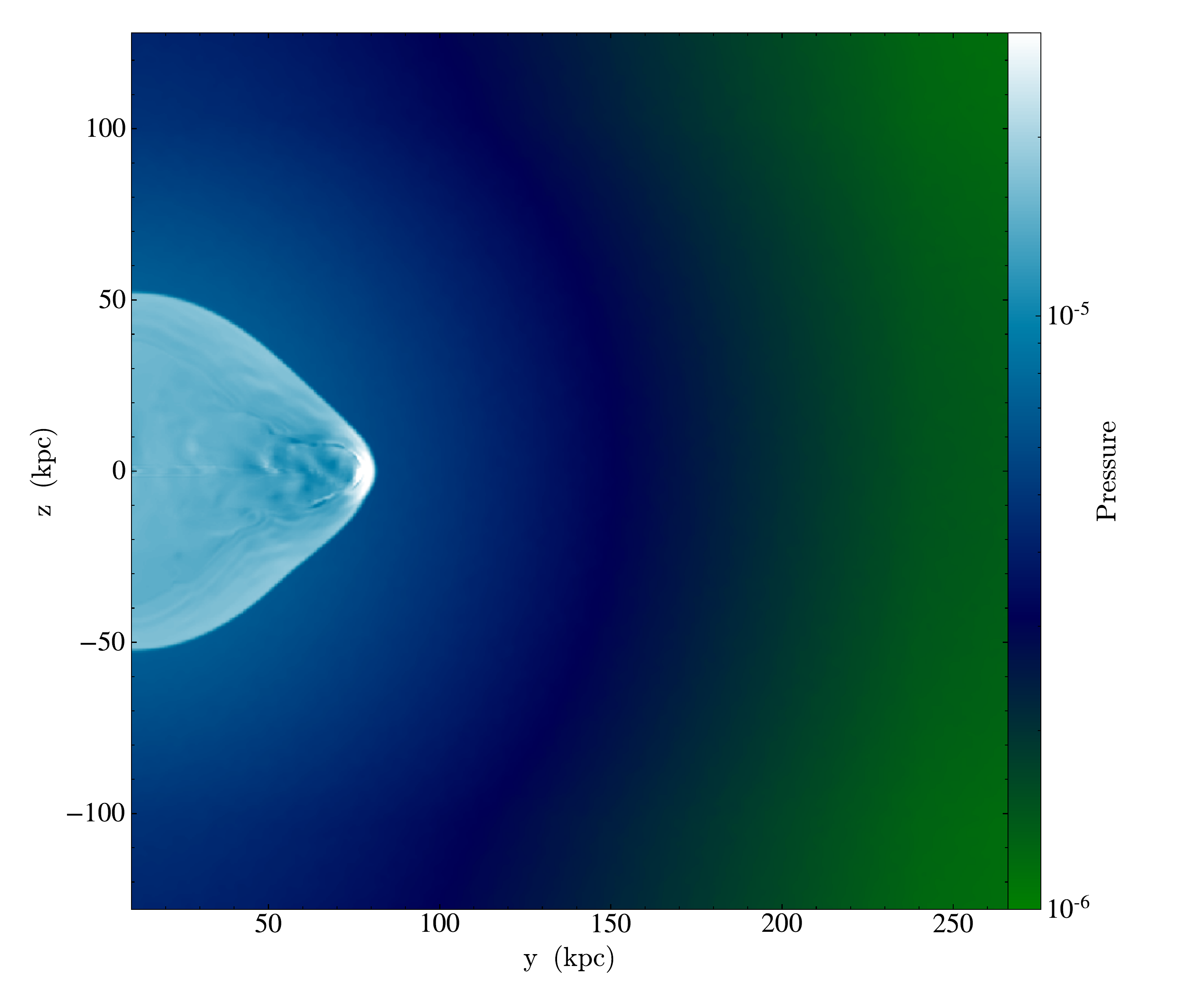}~    
    \includegraphics[width=0.33\linewidth]{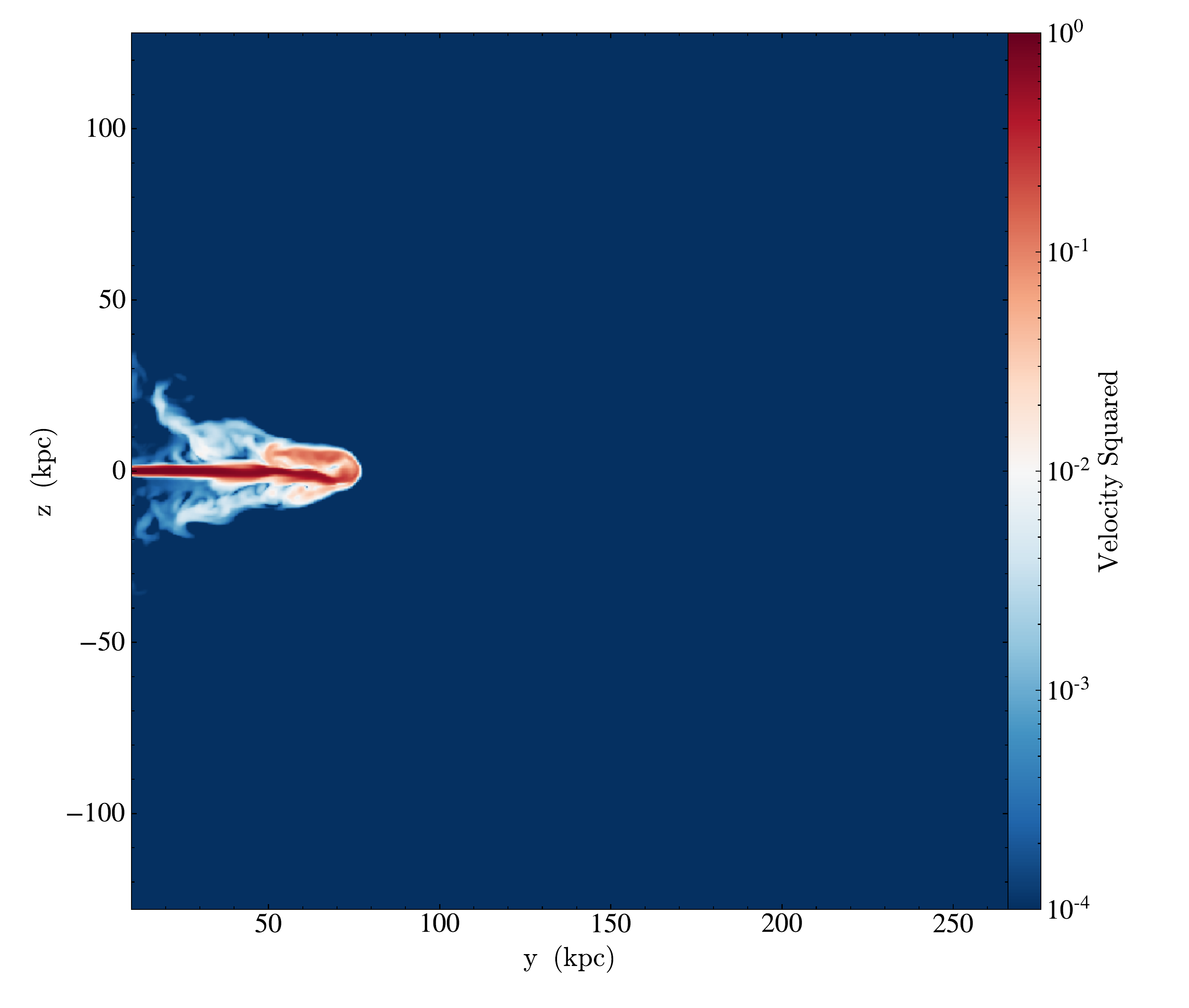}
        
	\includegraphics[width=0.33\linewidth]{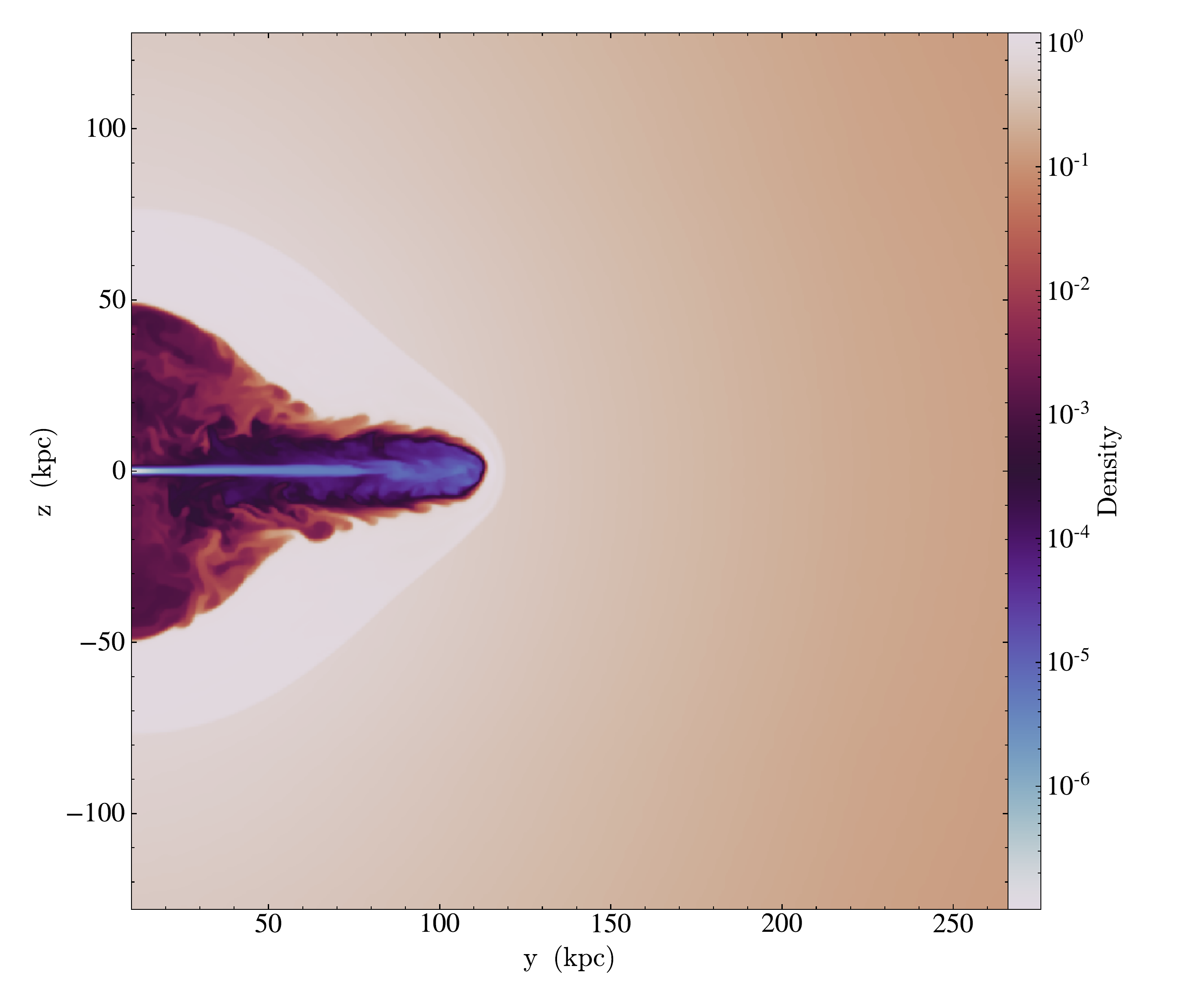}~    
    \includegraphics[width=0.33\linewidth]{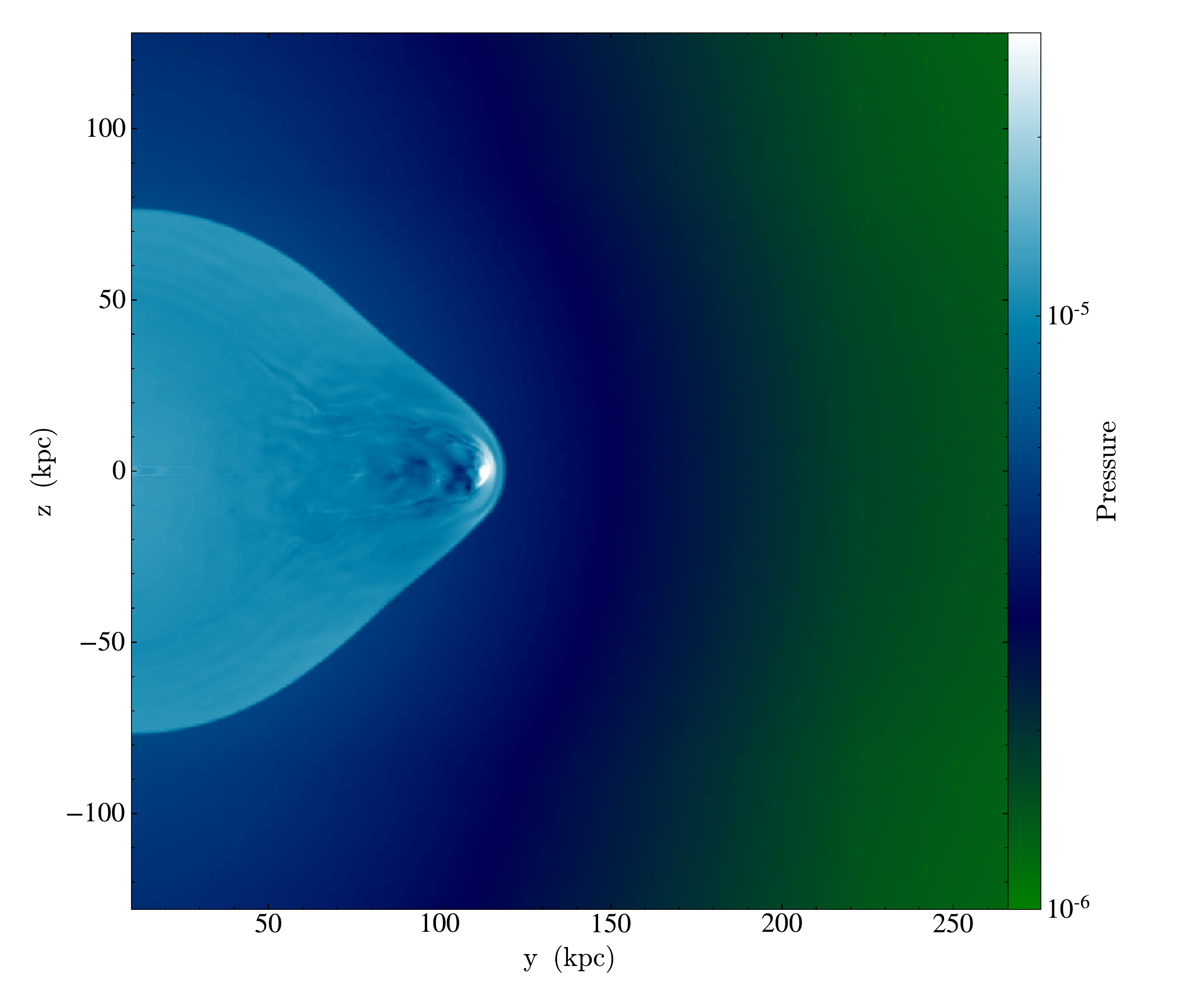}~    
    \includegraphics[width=0.33\linewidth]{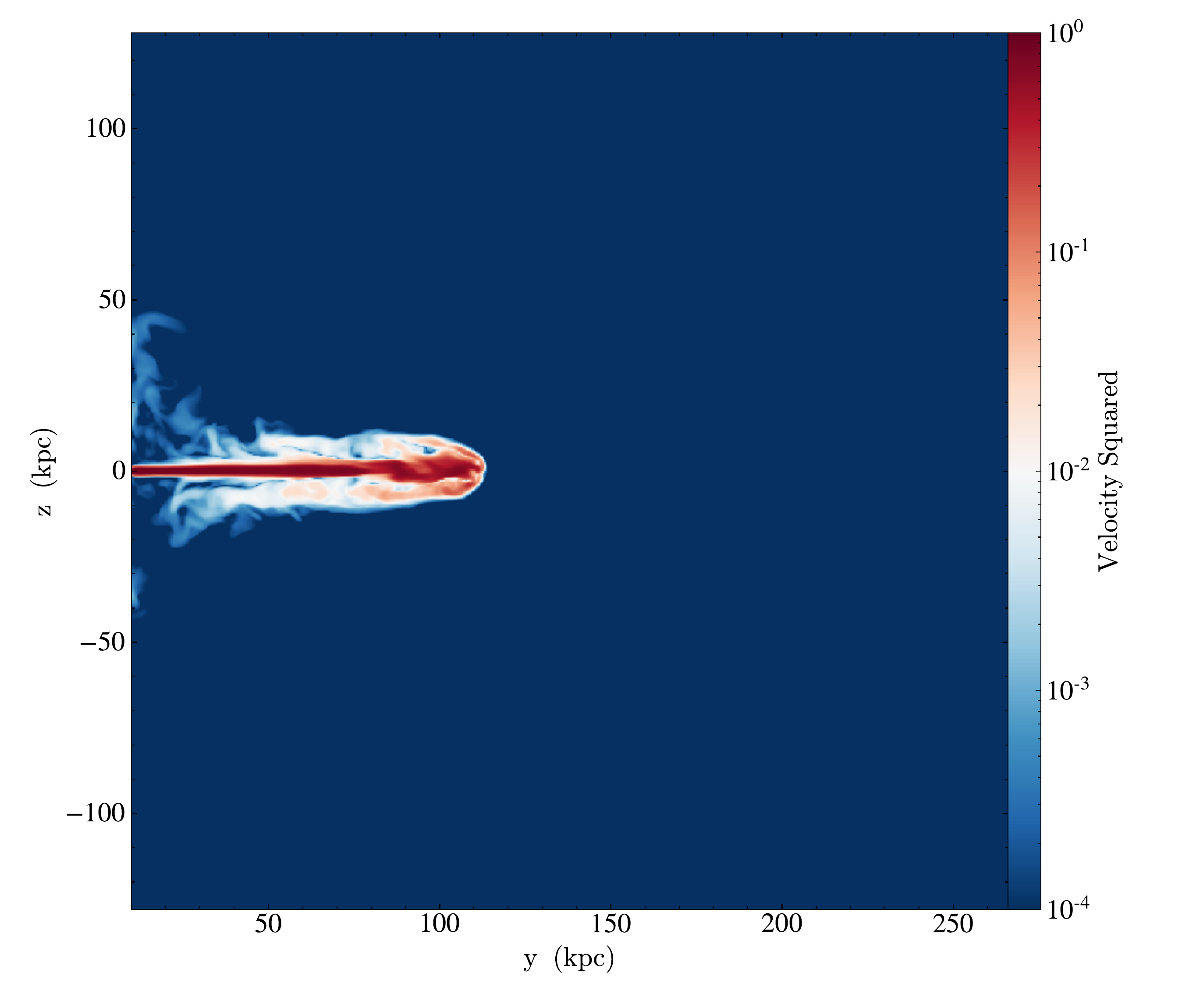}

	\includegraphics[width=0.33\linewidth]{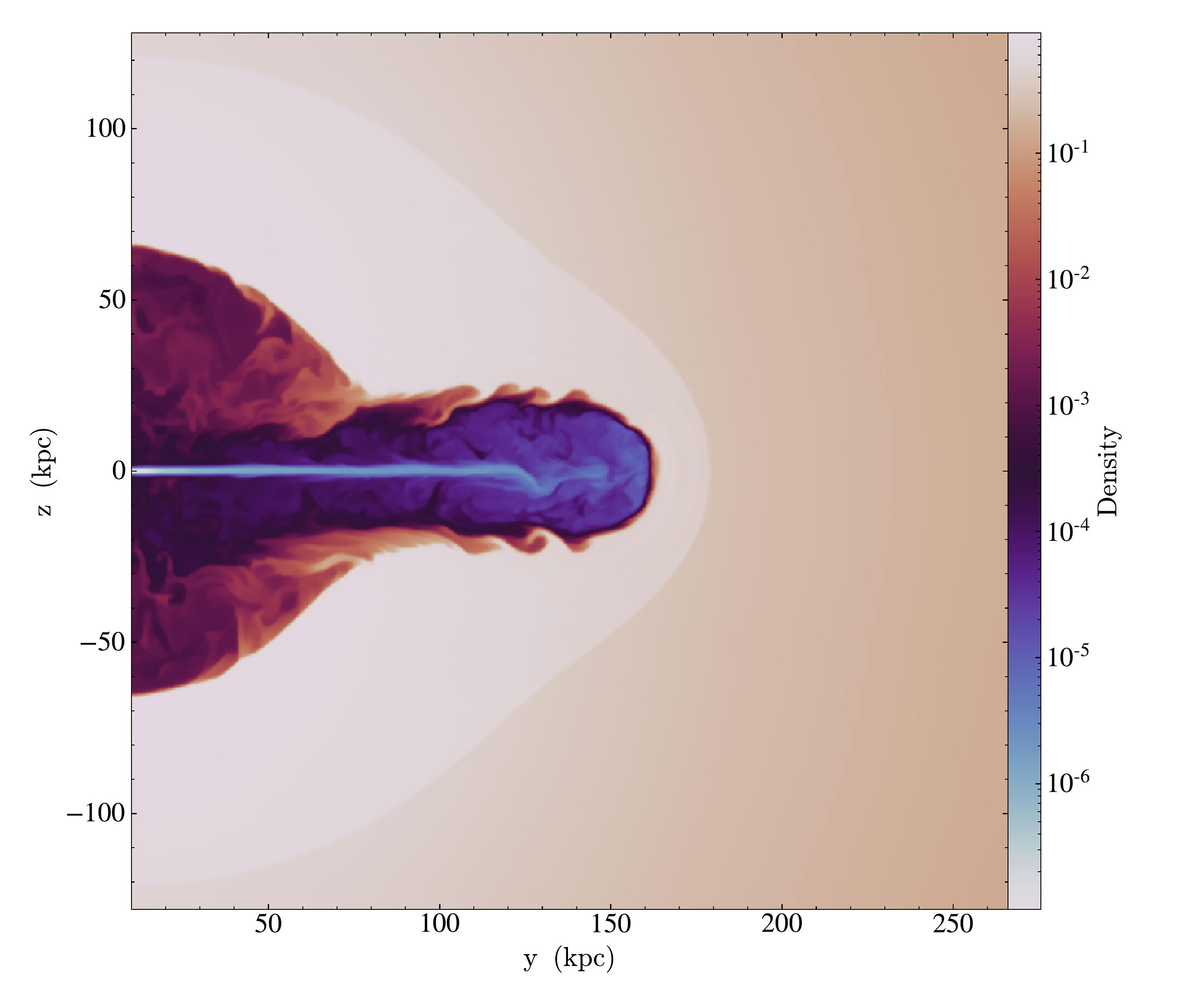}~    
    \includegraphics[width=0.33\linewidth]{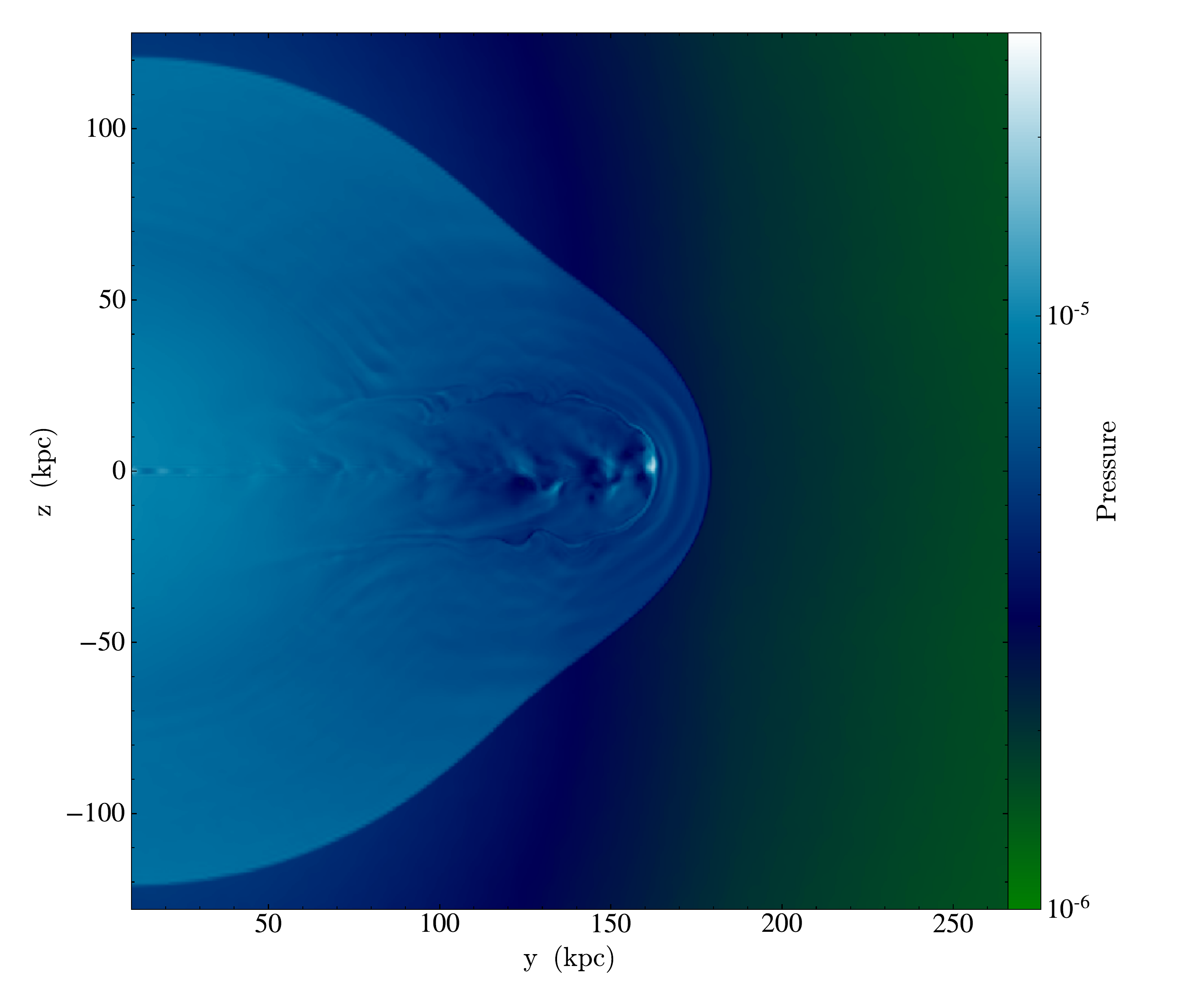}~    
    \includegraphics[width=0.33\linewidth]{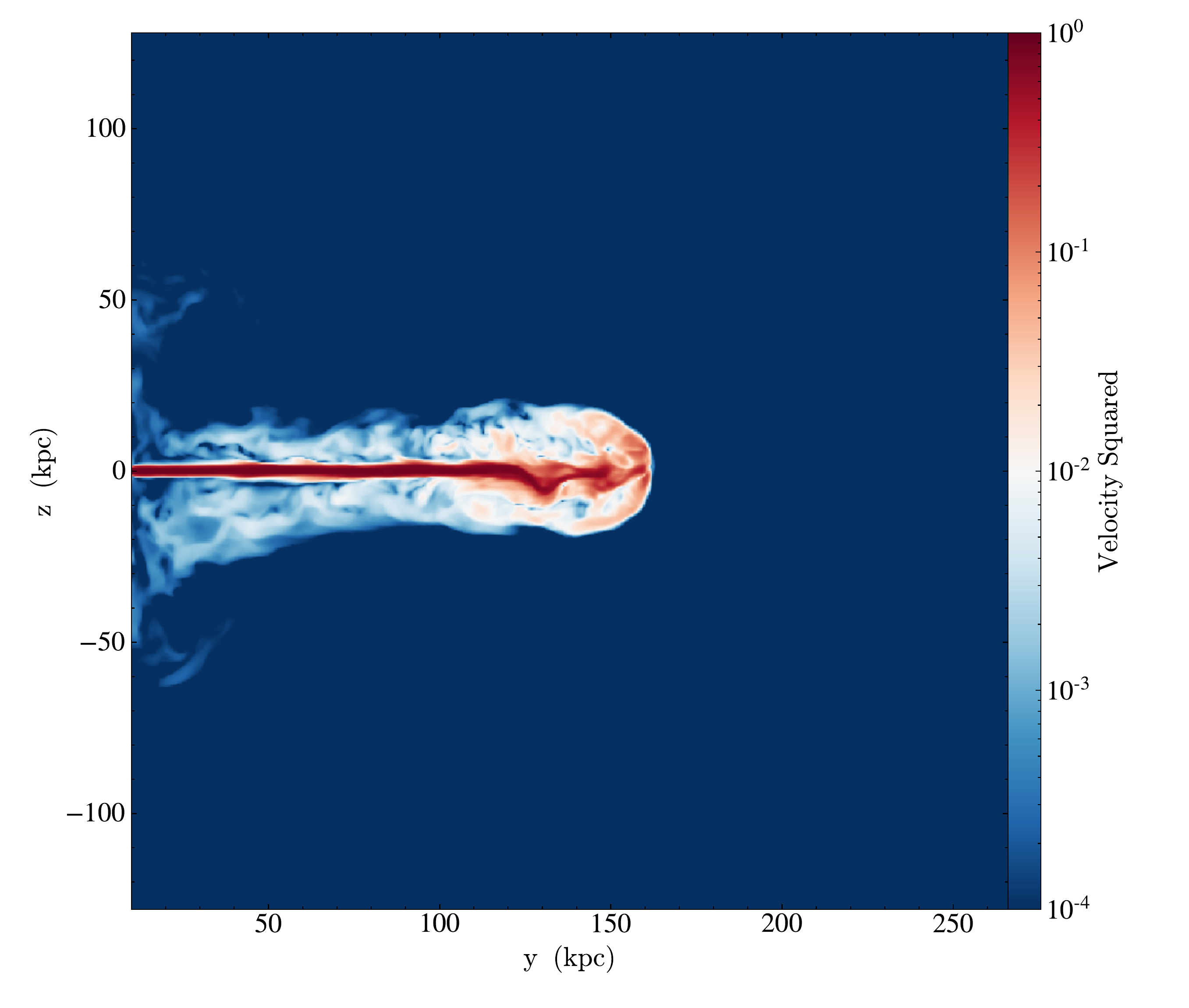}

	\includegraphics[width=0.33\linewidth]{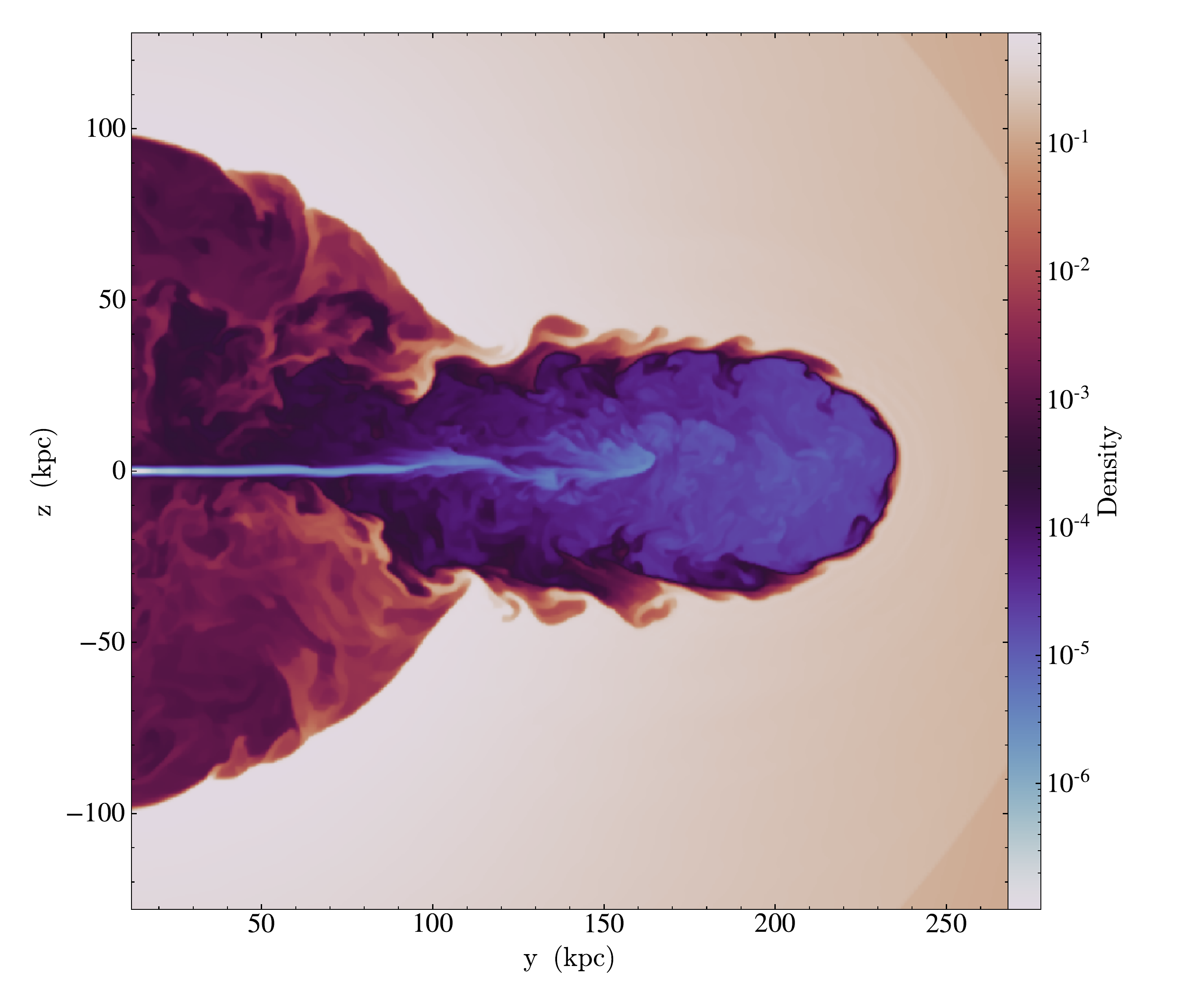}~    
    \includegraphics[width=0.33\linewidth]{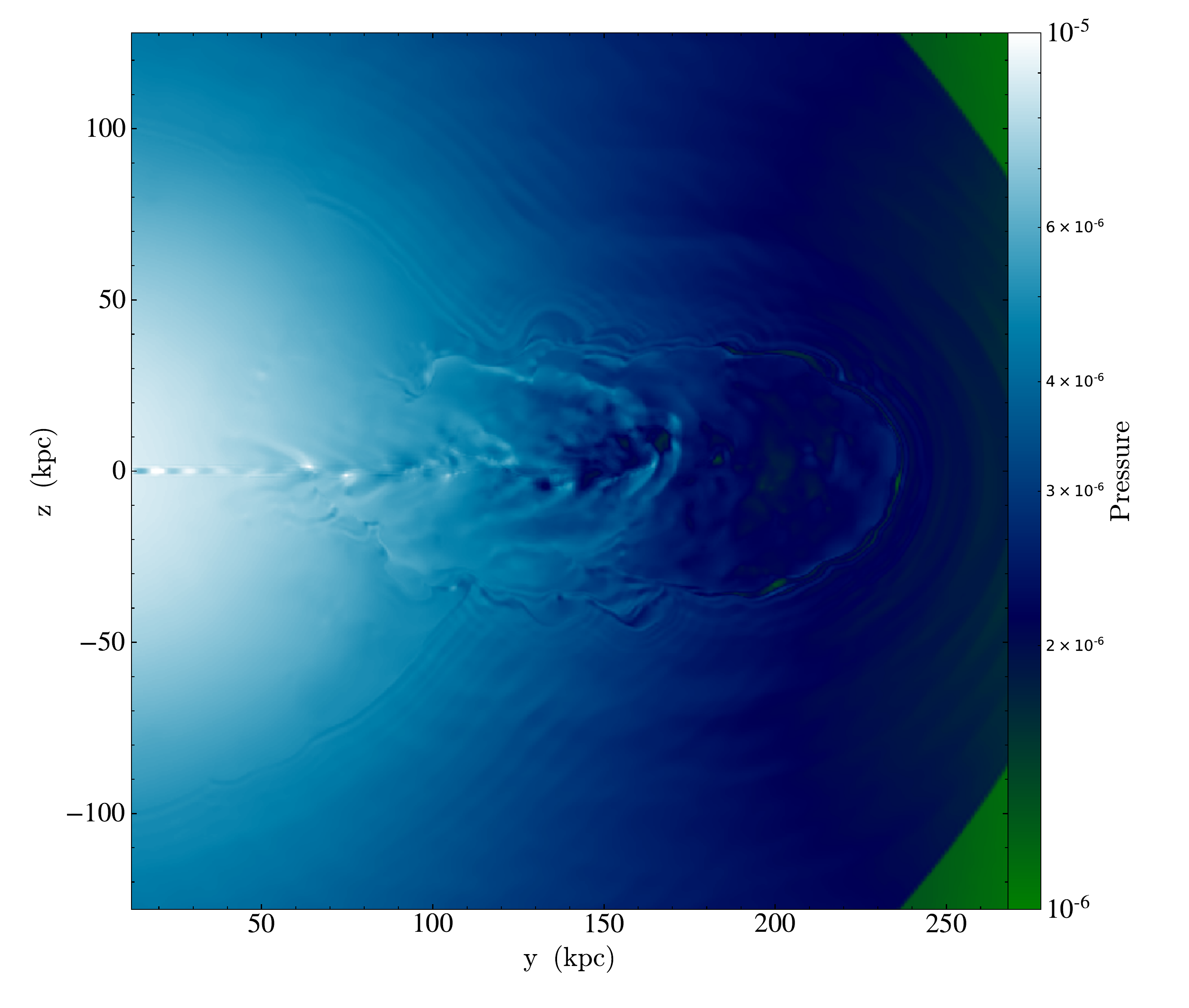}~    
    \includegraphics[width=0.33\linewidth]{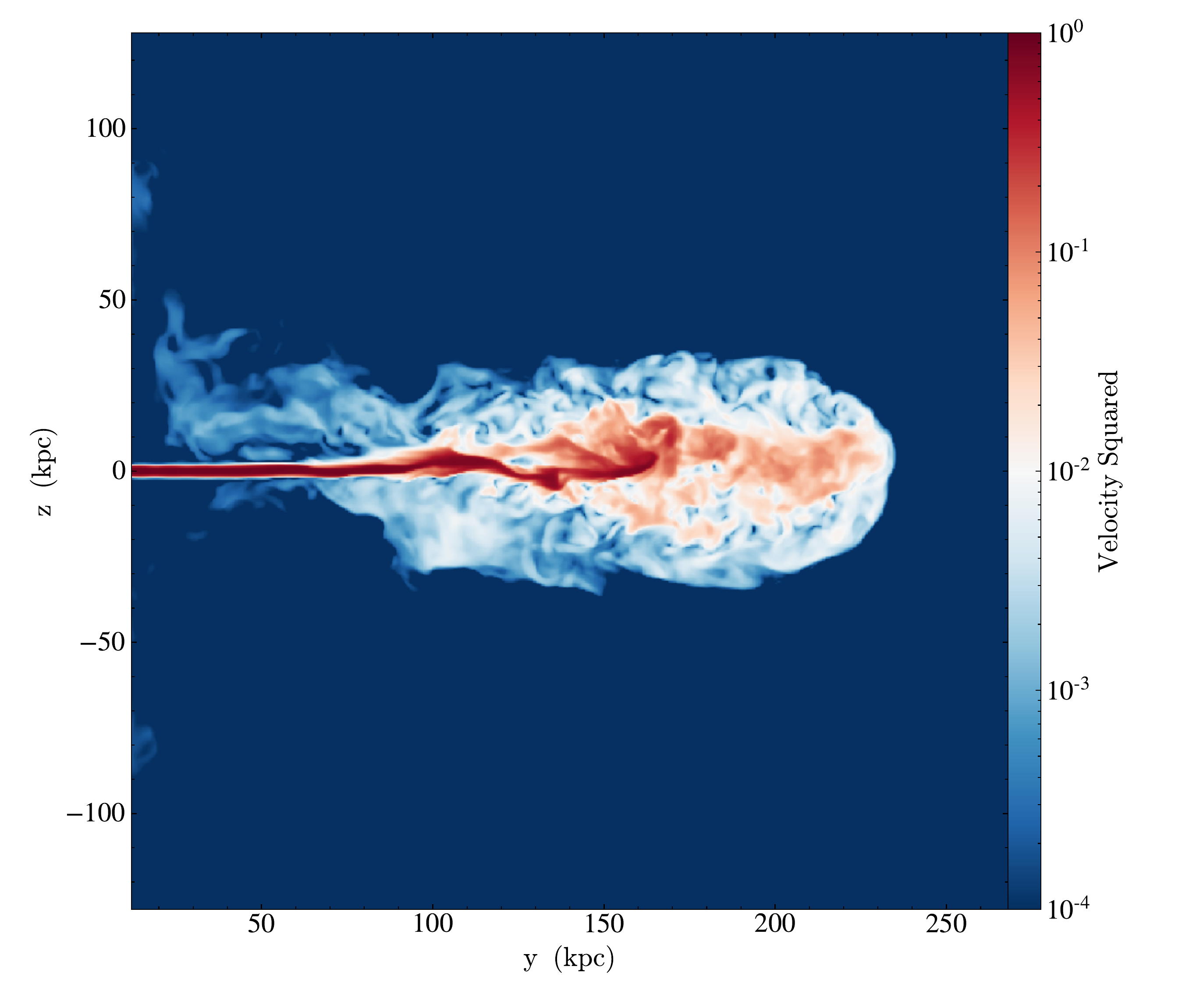}

    \caption{Rest-mass density (left panels), pressure (central panels) and velocity module squared (right panels) cuts for the internal energy dominated jet at times $21$ (first row), $36$ (second), $67$ (third), and $152 \,{\rm Myr}$ (fourth). 
    %The units are code units ($\rho_a$ for density, $\rho_a\,c^2$ for pressure, and $c$ for velocity).
   Units are: $\rho_a$ for density, $\rho_a\,c^2$ for pressure, and $c$ for velocity.}
  \label{fig:internal}
\end{figure*}

\subsection{Internal energy dominated jet}

\begin{figure*}%
	\includegraphics[width=\textwidth]{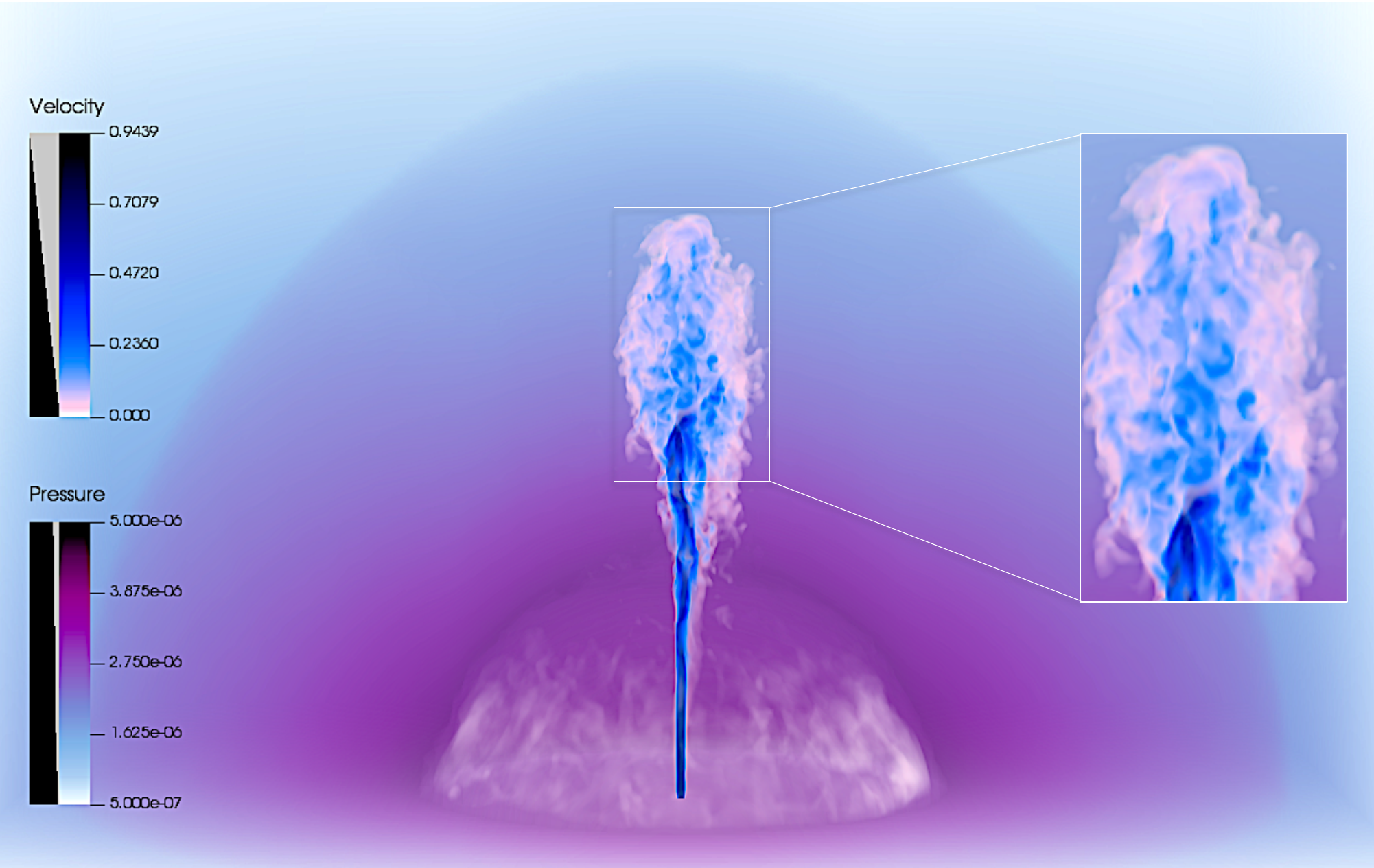}
    \caption{Pressure (purple scale) and velocity (blue scale, only positive values) rendering of the numerical box, highlighting the ambient medium and bow-shock, and the jet flow, respectively, at $t\simeq 152\,{\rm Myr}$. The spatial scale covers $\simeq 300\,{\rm kpc}$ along the vertical axis and $512\,{\rm kpc}$ along the horizontal axes. The inset shows a zoom of the positive axial velocity in the lobe. 
    %The units are code units ($\rho_a\,c^2$ for pressure, and $c$ for velocity).
    Units are: $\rho_a\,c^2$ for pressure, and $c$ for velocity.}
  \label{velpres}
\end{figure*}

Figure~\ref{fig:internal} shows cuts of rest-mass density, pressure and velocity module squared at different times along the simulation. Although along the initial stages of the evolution (two upper rows) the structure generated by the jet resembles that seen for case 3 in Fig.~\ref{fig:kinetic}, once the jet head reaches distances over $100$~kpc, the terminal shock detaches from the bow-shock and starts feeding a lobe with low-density, high-energy density, shocked gas. This hot gas drives the expansion of the radio source by means of its high pressure. The high sound speed in this region isotropizes pressure and the bow shock becomes quasi-spherical (this can be better seen in Fig.~\ref{velpres}). Thus, the lobes also tend to achieve this shape, in contrast with the elongated lobes obtained in the kinetically dominated simulations.
%trend of the bow-shock towards a spherical shape }

Figure~\ref{velpres} displays renderings of pressure and axial velocity distributions at the end of this simulation, i.e., after $\sim 152\,{\rm Myr}$. At this time, the front shock is at $280\,{\rm kpc}$.\footnote{The advance velocity of the shock
ranges from $\sim 0.015\,c$ at the beginning of the simulation to $\sim 5\times10^{-3}\,c$ at the end.} The purple scale shows the rendering of pressure distribution in code units ($\rho_a\,c^2$, with $\rho_a$ the ISM density at injection, i.e., $1.67\times 10^{-26}\,{\rm g/cm^3}$). 
The image captures the mild shock around the jet, with an almost spherical shape, sweeping the smooth ambient medium distribution.
The simulation had to be stopped at this point to avoid the bow-shock crossing the grid boundary. %Therefore, the development of the simulated jet did not reach the scales of the model source, Hercules A, where each of both jets reach distances $\geq 250\,{\rm kpc}$.   

The blue scale in Fig.~\ref{velpres} shows the axial velocity distribution -limited to positive values-, which reveals the inner jet structure, the lobes, and the hot bubble of shocked jet material accumulated by the host galaxy gravitational potential.\footnote{The dynamical time scale estimated for the gravitational field, $t_d\sim1/\sqrt{G\,\rho_{\mathrm{DM}}}$ (with $G$ the gravitational constant and $\rho_{\mathrm{DM}}$ the central density of the dark matter halo deduced from the hydrostatic equilibrium condition),
is $\simeq 180\,{\rm Myr}$. Although the simulation time is slightly shorter ($\simeq 150\,{\rm Myr}$), 
%the backflow velocities accelerate the fall of the shocked gas and favour the spherical distribution observed
the blackflow velocities found in the lobes accelerate the fall of the shocked jet gas towards the galactic core and the formation of the observed spherical distribution \citep[see also][]{2014MNRAS.441.1488P}.} The jet shows a kinked morphology due to the development of KHI modes, excited by the helical motions set up as injection boundary conditions. The growth of these modes is faster in hot flows \citep[e.g.,][]{2005A&A...443..863P}, and they trigger oscillations in the physical variables of the jet \citep[e.g.,][]{2000ApJ...533..176H}. Indeed, the image shows the oscillations in axial jet velocity, which end up with the disruption of the flow at $\sim 70\,{\rm kpc}$ from injection, i.e., $\sim 80\,{\rm kpc}$ from the galactic nucleus. 

The simulated jet does not show expansions and recollimations before the kinks become non-linear, in agreement with the observations of the jets in Hercules~A previous to disruption \citep[][]{2013ApJ...771...38O}, and in contrast to other simulations of low-power, hot relativistic jets \citep{2007MNRAS.382..526P} or non-relativistic, low Mach number jets \citep{2002ApJ...579..176S}. It is worth to mention that the Mach disks reported to be responsible for jet disruption in those works are common in axisymmetric simulations because of the imposed symmetry. These kind of structures are hardly found in three-dimensional simulations. The reason is the high cocoon pressure, which keeps the jet collimated and prevents its fast expansion. As shown by \cite{2017MNRAS.471L.120P}, the lobe pressure correlates with the flux of available (i.e., internal plus kinetic) energy transported by the jet. In the case of the present simulation, the internal energy of the jet at injection, together with the large scale density/pressure core helps to keep the jet collimated.

The disruption point of the jet stands in the region $100-150\,{\rm kpc}$ since the jet head reaches $\sim 100\,{\rm kpc}$ (at $t \approx 36$ My), and oscillates back and forth throughout the rest of the simulation ($t \approx 152$ My) (note that disruption is farther from injection in the third row of Figure~\ref{fig:internal} than in the fourth, $\simeq 80\,{\rm Myr}$ later). This fact indicates that disruption is caused by the development of helical instabilities and that the precise point at which it takes place may change slightly due to changes in local conditions. However, 
as far as conditions remain unchanged at injection, it is difficult to foresee a dramatic change in the jet stability conditions.
%, i.e., a significant approach of the terminal shock towards the bow shock. 
Therefore, we can conclude that the final snapshot of the simulation is not a transitory phase. Injection conditions should be changed in order to see changes in the whole jet/lobe structure, which would, furthermore, require times of the order of the simulated one $\sim 100\,{\rm Myr}$.

\begin{figure*}%
    
    \includegraphics[width=\linewidth]{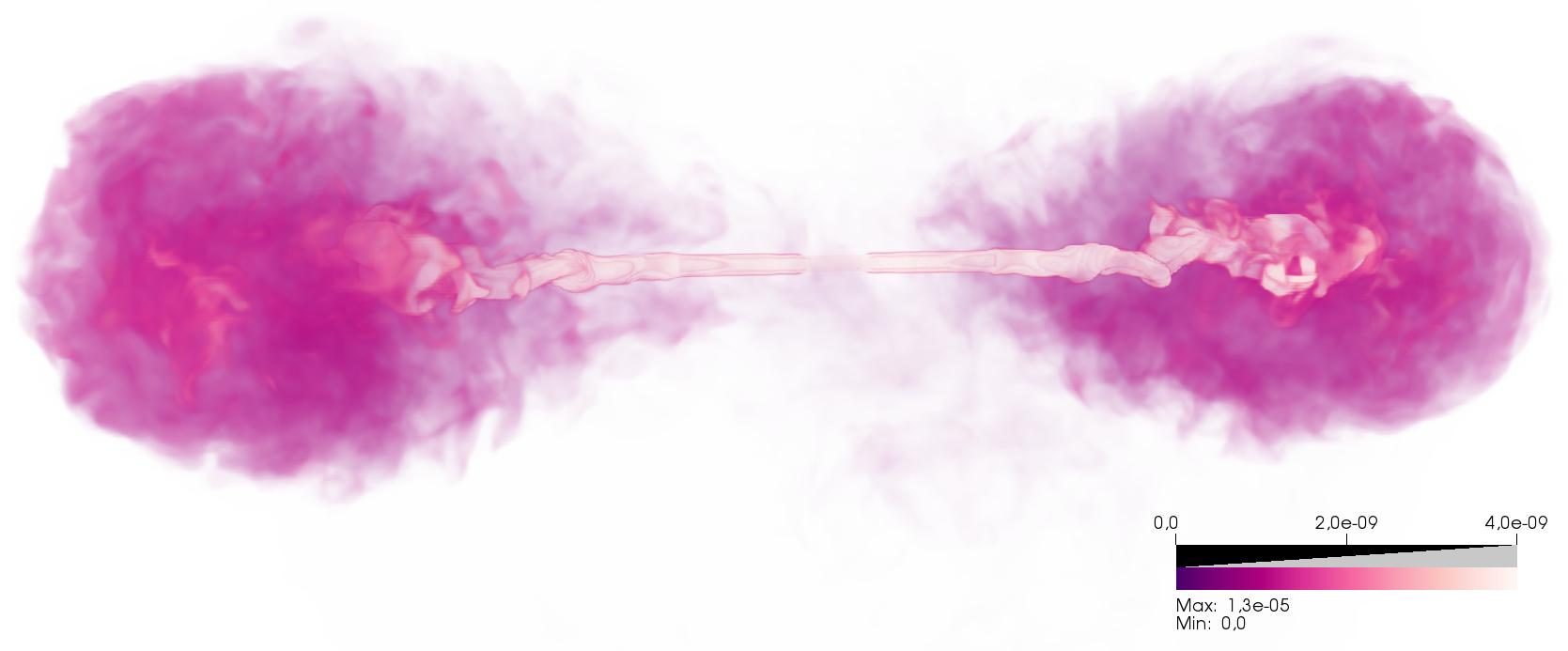}
    
%    \vspace{0.1cm}
%    \includegraphics[width=\linewidth]{eps_mod.png}  
    %\includegraphics[width=0.9\linewidth]{eps3_2.png}
	%\includegraphics[width=0.9\linewidth]{eps2.png}
    %\includegraphics[width=0.9\linewidth]{eps.png}
    %\includegraphics[width=\linewidth]{eps4.png}
    \caption{Pressure weighted by tracer rendering of the numerical box. The simulated jet has been mirrored to emulate a double sided jet and show a different perspective of the source.} %(top) and specific internal energy (bottom)}
  \label{eps}
\end{figure*}

Figure~\ref{eps} shows a rendering of pressure (weighted by tracer) (in units of $\rho_a\,c^2$) to be compared with Fig.~\ref{fig:kinetic2} for kinetically dominated jets. The image is a mirrored composition of the simulated jet, which has been rotated to a viewing angle of $45^\circ$ counterclockwise (around a vertical axis) to show two perspectives of the three-dimensional structure. There is an evident difference with the kinetically dominated jets in terms of lobe shape, which is more spherical in this simulation. The image reveals the kink of the jet before disruption inside the lobes, and their turbulent internal appearance. The kinks and disruption region show high values of internal energy as a consequence of the dissipation of kinetic energy. 

\begin{figure*}%
	\includegraphics[width=0.5\textwidth]{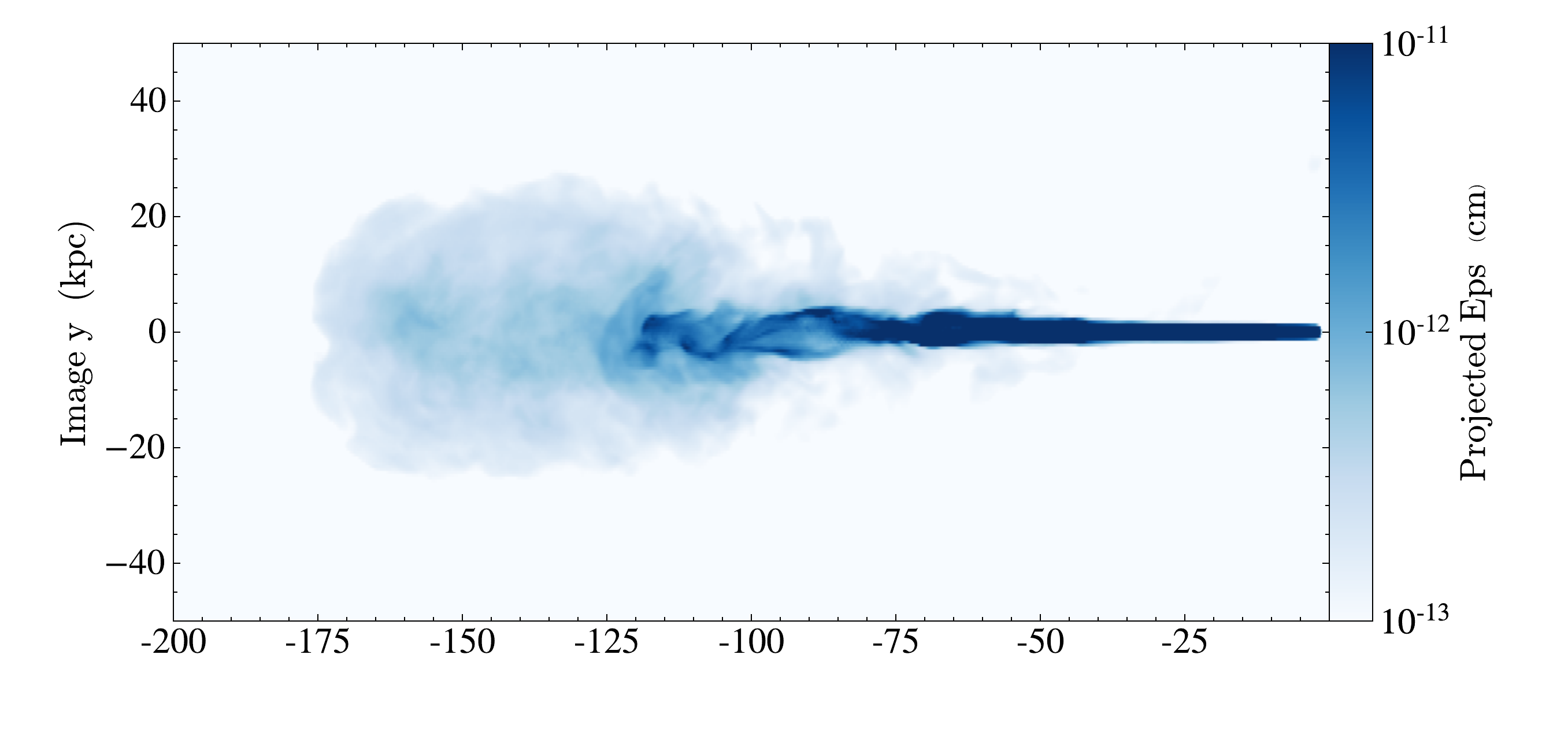}~
     \includegraphics[width=0.5\textwidth]{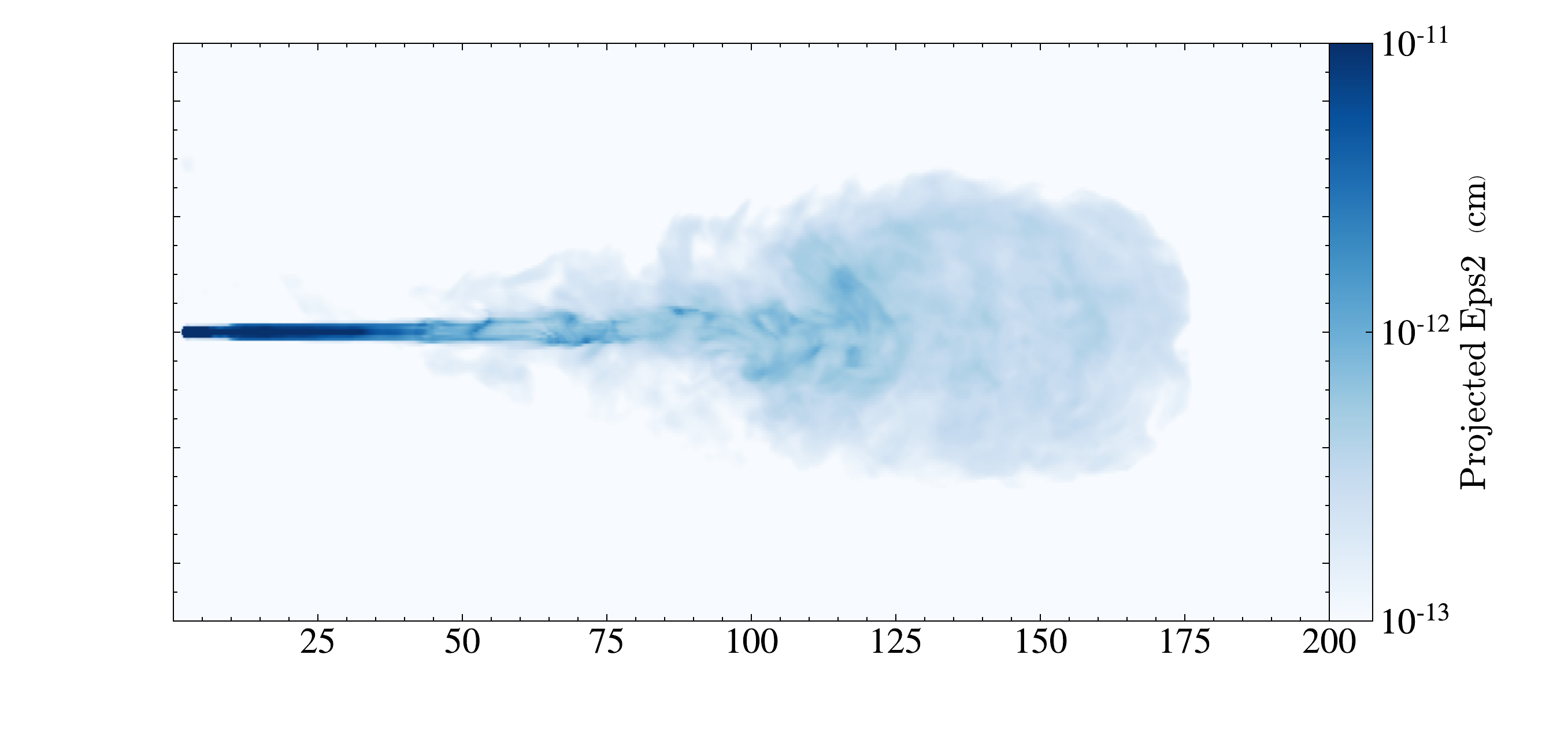}
	\includegraphics[width=0.5\textwidth]{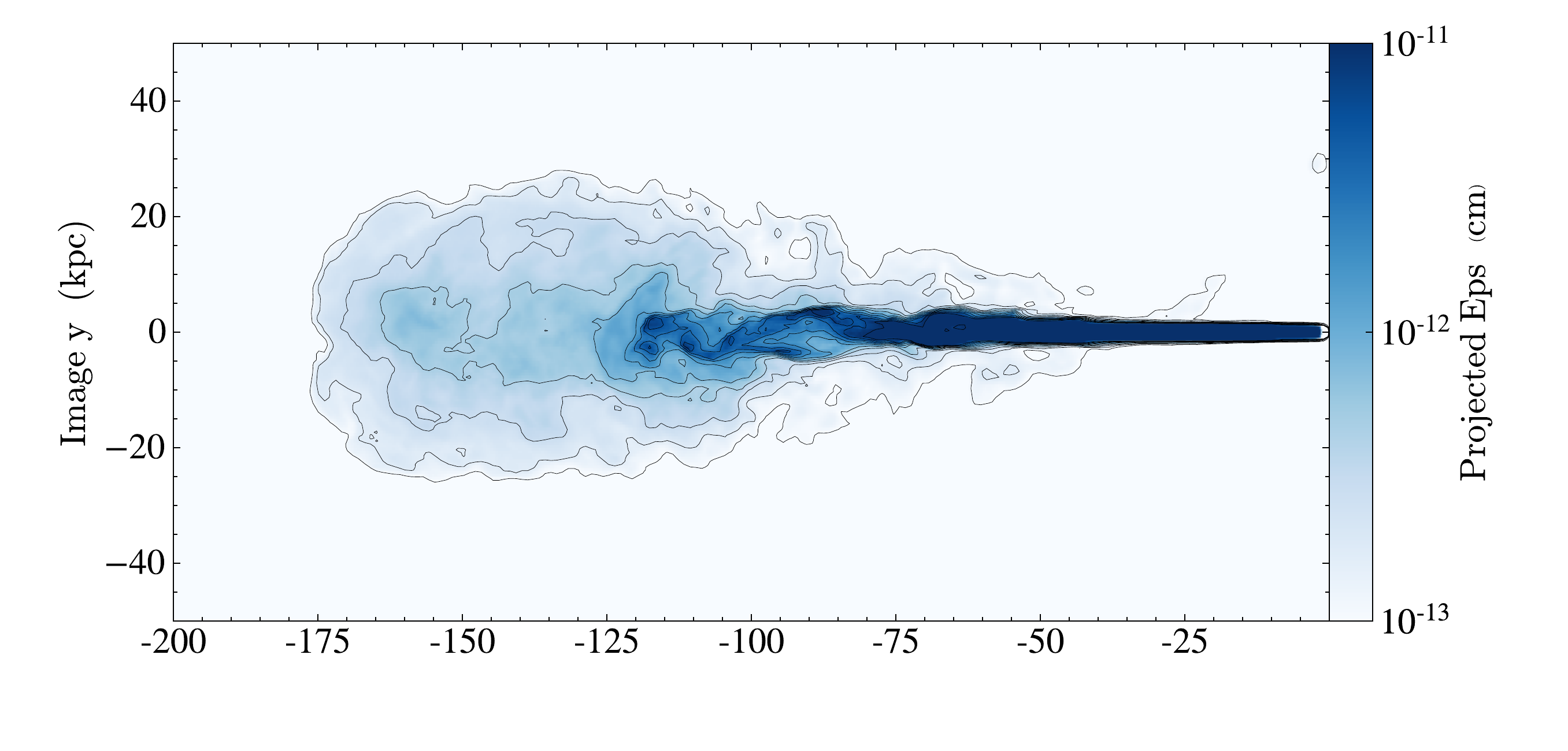}~
     \includegraphics[width=0.5\textwidth]{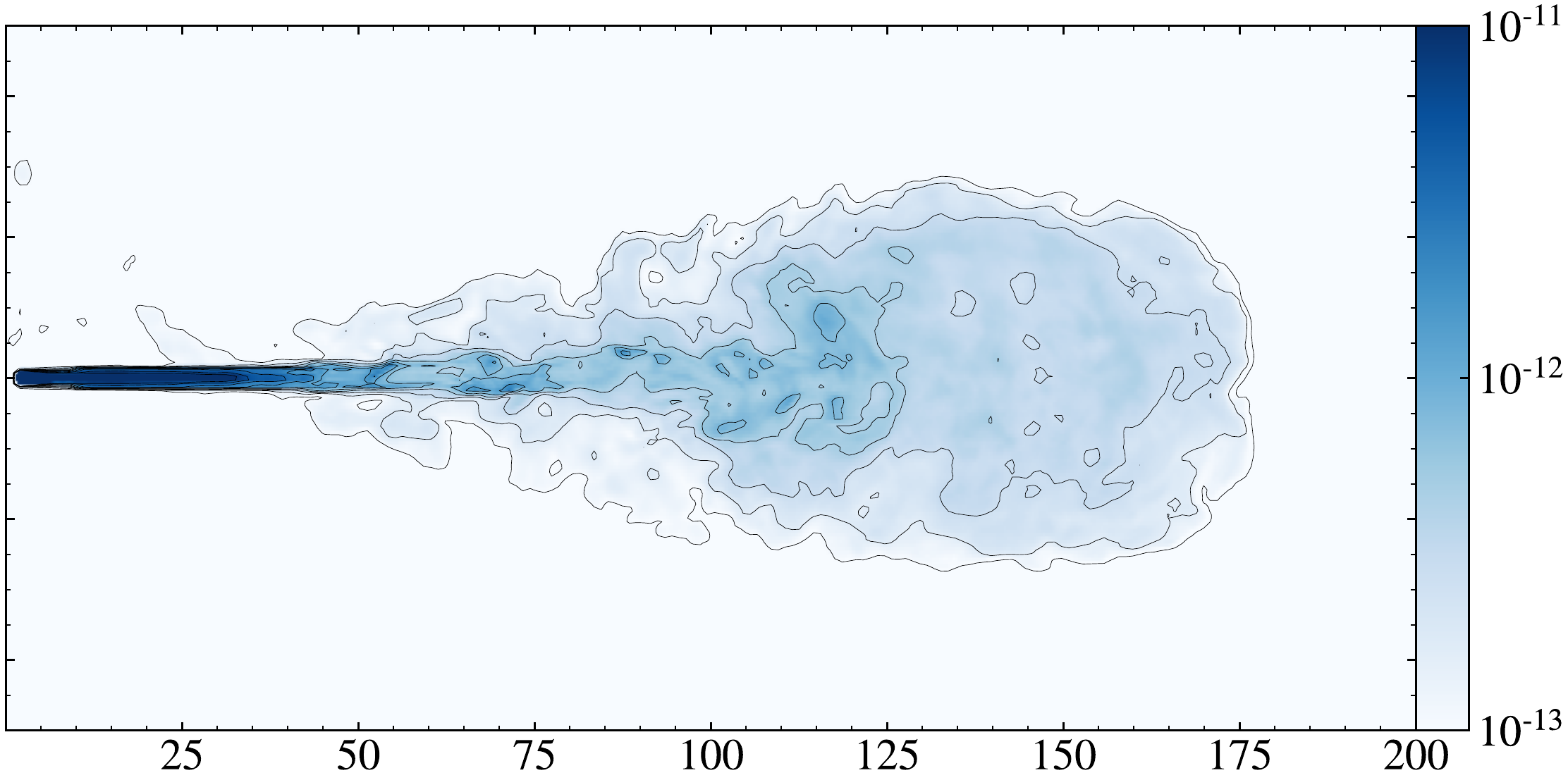}
    \caption{Pseudo-synchrotron emissivity images, in arbitrary units, for the jet at its last snapshot, projected at $50^\circ$ (left) and $130^\circ$ (right) viewing angles.}
  \label{fig:sync}
\end{figure*}

Figure~\ref{fig:sync} shows projected pseudosynchrotron emissivity images for the jet at its last snapshot at $50^\circ$ (eastern lobe; left) and $130^\circ$ (western lobe; right) viewing angles, in arbitrary units, following the results from radio and X-ray observations \citep[e.g.,][]{2003MNRAS.342..399G}. The expression used to compute the emissivity at a given frequency is the same as that used by \citet{2003ApJ...597..798H} \citep[see also][]{1989ApJ...342..700C}: 
\begin{equation}
\epsilon_\nu \,\propto\,n^{1-2\alpha}\,p^{2\alpha}\,(B\,\sin\theta_B)^{1+\alpha}\,D^{2+\alpha},
\end{equation}
where $n$ is the lepton number density, $p$ is pressure, $B$ is the assumed magnetic field strength in the comoving frame, taken to be proportional to $\sqrt{p}$, $\theta_B$ is the angle between the field lines and the viewing angle (also in the comoving frame) assumed to be a constant, $D$ is the Doppler factor, and $\alpha$ is the spectral index (defined as $S_\nu \propto \nu^{-\alpha}$). 
The spectral index has been defined as a linear function of the jet gas tracer $f$, $\alpha\,=\,0.6\,+\,0.9\,(1-f)$, to obtain values from 0.6 (in the jets) to 1.5 \citep[in the lobes;][]{2003MNRAS.342..399G,2005MNRAS.358.1061G,2022A&A...658A...5T}.
The Doppler factor has been computed using only the axial velocity component. The image shows that the simulated lobes extend up to $\simeq 175\,{\rm kpc}$ in projection, whereas the jets in Hercules~A reach $\simeq 260\,{\rm kpc}$, which explains that the lobes in Hercules~A are more inflated. The advance velocity of the radio structure along the jet axis falls rapidly after disruption and as the jet expands. Thus, it is reasonable to expect a slow evolution of the lobes towards a spherical shape within the remaining $\simeq 80-90\,{\rm kpc}$ that the simulation would need to reach the 
actual size of Hercules~A.\footnote{Reaching these scales would imply an amount computing time far beyond that used for this work.}
%In the image, the width ratio between the jet and the lobe is $\sim 0.04$, quite close to the VLA image for the eastern lobe, if we consider the jet width prior to disruption ($\sim 0.025$). Therefore, if the jet width does not change while the lobe keeps expanding, the observed ratio could be easily achieved.

Despite the simplifications, the images show interesting features, namely, bright jets that develop helical patterns and disrupt into the lobes, and arcs (e.g., at $\simeq 120\,{\rm kpc}$ in the western jet) that are produced by the disruption of the jet when the helical instability achieves large amplitude and forces parts of the jet to generate shocks into its environment. This effect was also observed in numerical simulations that tackled the development of KHI in relativistic flows \citep{2005A&A...443..863P}. The possibility that the arcs could be helical features observed in projection was actually suggested by \citet{2003MNRAS.342..399G}. Keeping in mind that our simulation did not reach the extension of Hercules~A, perhaps the simulation shows the beginning of the formation of those arcs as instability-generated shocks. Therefore, it would be necessary to continue the simulation to make a better comparison. Interestingly, the hints of arc-like structures are mainly observed in the right panel (western jet). All other things being equal (these images are produced from a single simulated jet), this difference is caused by the viewing angle: Doppler boosting makes the eastern jet brighter and reduces the contrast that allows the arc-shaped structure to be enhanced in the western jet.

Finally, we can also compare the lobe pressure at the end of the simulation with the estimated one in \citet{2004MNRAS.350..865G} for the lobes in Hercules~A, as a test for the validity of our approach. In the simulations, we find $P_l\simeq 10^{-10} \,{\rm dyn/cm^2}$, as compared to $P_l\simeq 2-3\times 10^{-11} \,{\rm dyn/cm^2}$ from observations. Taking into account the aforementioned difference in size and that the cocoon pressure evolves with distance as $d^{-0.9}$ \citep[a fit obtained as explained in][]{2019MNRAS.482.3718P}, the difference 
in lobe pressures could be reduced from a factor 3-5 to a factor 2-3. This discrepancy is possibly due to the large injection power \citep[see][]{2017MNRAS.471L.120P} used in the simulation with respect to the expected current jet power in Hercules~A (see the discussion).

\section{Discussion} \label{sec:disc}

\subsection{Global morphology}

The sphericity of the lobes (accentuated by the perspective) is a sign of a pressure-driven isotropic expansion of the lobes. It has been suggested that the lobes are fed by a possibly intermittent jet \citep[e.g.,][]{2022A&A...658A...5T} with a periodicity of the order of a factor $10^5{\rm yr}$ \citep[within a long activity burst of $\simeq 60\,{\rm Myr}$,][]{Nulsen2005}. The jet shows continuous emission from the source to the disruption point, implying that the activity periods must be long enough to allow the plasma to reach $\sim 100\,{\rm kpc}$. According to the authors, for an estimated velocity of $0.8\,c$, this implies a minimum activity time of $4\times10^5\,{\rm yr}$. 
If this is correct, our simulations (which do not consider such a short periodicity) would then represent the continuous injection of the mean properties of the jet, i.e., the long activity burst. Nevertheless, this is irrelevant for the main point revealed in this paper: the global morphology can be recovered by jets dominated by internal energy flux, and not by those kinetically dominated. It is possible, though, that such a fast periodicity triggers internal shocks in the jets that dissipate kinetic energy along the jet channel, heating the flow as it evolves, relaxing the requirement of a very large internal energy from injection that we have considered in our simulation.

The idea behind the setup of the simulations was actually to provide a plausible explanation to the overall morphology of the radio galaxy Hercules A, which does not fit into the classical FRI/FRII classification. The initial hypothesis was inspired by the sphericity of the lobes plus the fact that the lobe expansion takes place beyond the density/pressure fall of the ambient medium profiles. Also the results obtained in a series of numerical and theoretical works on jet evolution \citep{2011ApJ...743...42P,2014MNRAS.445.1462P,2017MNRAS.471L.120P} showed that the morphology of the lobes depends on whether the jet energy flux is dominated by the kinetic energy (mass of the particles) or by the internal energy. In the first case the inertia of the flow tend to give rise to elongated structures. In the second case, the expansion of the lobes, driven by their high pressure, is isotropic. Furthermore, it was concluded that this second option is limited to relativistically hot jets, in which the internal energy budget is large enough to dominate, or at the least, compete, with kinetic energy flux. Therefore, this effect cannot be studied with non-relativistic simulations \citep{2002ApJ...579..176S}. 

The setup was thought to keep jet collimation by means of environment pressure, at the same time as developing unstable (helical) KHI modes that trigger disruption; all these requirements are fulfilled by a (slowly expanding) relativistically hot jet. Those properties produce a high pressure lobe, as expected from previous theoretical work \citep{2017MNRAS.471L.120P}, and keep the jet confined. In addition, the lack of significant jet expansion \citep[][showed that jet expansion decelerates the growth of instability amplitudes]{1982ApJ...257..509H} together with the high internal energy of the flow \citep{2005A&A...443..863P} cause a relatively faster development of the KHI modes. As a result, the simulated jet evolution shows a global jet and lobe morphology that are similar to the observed eastern jet/lobe system, in terms of development of unstable kink patterns, jet disruption within the lobes, the lack of hotspots, and  a more realistic jet-to-lobe 
width ratio \citep[$\sim 2.7\times10^{-2}$ in the source,][versus $\sim 4.5\times10^{-2}$ in the simulation, see Fig.~\ref{fig:sync}]{2003MNRAS.342..399G} between the jet and the lobe at the current simulated size. 
All in all, the basic morphological and dynamical properties of the jets and lobes of Hercules~A (kink unstable jets, quasi-spherical lobes, lack of hotspots, right jet-to-lobe width ratio) are reproduced by a relativistic jet with high internal energy.

\citet{2016A&A...596A..12M} have shown that low power jets ($L_j \leq 10^{43}\,{\rm erg/s}$) can result in a similar morphology, mainly due to the dissipation of kinetic energy at disruption. \citet{2021MNRAS.508.5239Y} have also obtained similar structures for jets injected with an opening angle into the grid. However, the jet-lobe structure is not expected for more powerful jets with small opening angles, like the one we simulate.

Taking into account that kinetically dominated jets generate elongated lobes \citep[see][for a comparison between large scale morphology of cold and hot flows]{2014MNRAS.445.1462P}, our work shows evidence favouring that the jets in Hercules A are not kinetically dominated at large scales, but by internal energy. 
As noted in the Introduction, we have not included magnetic fields in our simulations. However, the isotropic expansion of the lobes requires the contribution of the magnetic field to be mainly in the form of pressure, i.e., requires the magnetic field to be disordered. \citet{2003MNRAS.342..399G} showed that the magnetic field is aligned with the lobe surfaces \citep[see also][]{2011MNRAS.413.2525G}. However, the lobes seem to be less polarized towards the central regions, which could be an indication of a disordered field, consistently with the turbulent character of the flow within the lobes. The contribution of such a magnetic field to the simulations would have not changed the obtained morphology. However, the role of a dynamically relevant toroidal field in the jet remains a plausible option to explain the development of the kink instability that finally disrupts the jet,\footnote{The presence of a toroidal field could, on the one hand, relax the high values of the internal energy flux in the jet and, on the other, trigger the development of current-driven instability modes.} which should be explored by means of RMHD simulations, and will be left for future work. This, again, would not change our conclusion, since the field would become disordered beyond the disruption point, which is supported by the turbulent dynamics of the region and the relative depolarization observed at the lobe central regions \citep[][]{2003MNRAS.342..399G}.

%\begin{figure}%
%	\includegraphics[width=\columnwidth]{density.png}
%    \caption{ISM/IGM density profiles in simulations presented in \citet{2014MNRAS.445.1462P,2019MNRAS.482.3718P} –red dashed–, \citet{2022MNRAS.510.2084P} –dotted blue–, and this work –solid blue. At the simulated scales, the medium is considered isothermal. Thus, these profiles are equal for the pressure distributions.}
%  \label{density}
%\end{figure}

A last point about the jets in Hercules~A concerns the origin of structures seen inside the western lobe and whether they are associated with the disruption of the jet. Comparing with pseudo-syncrhotron images from their numerical simulations, \cite{2002ApJ...579..176S} interpret these structures as rings associated to nearly annular shocks propagating through the backflow surrounding the jet. However, on the one hand, the ring-like structures seen in the pseudo-synchrotron images are probably an artifact produced by the imposed axisymmetry, as also observed in \citet{2002MNRAS.331..615S} for completely different simulations and setups. On the other hand, the detailed imaging of the source \citep[e.g.,][]{2003MNRAS.342..399G,2022A&A...658A...5T} reveals that the jet propagates inside the first expansion region \citep[structure E in][]{2003MNRAS.342..399G}. Moreover, if interpreted as a jet expansion, it would imply a sudden change in jet-to-lobe pressure ratio that is difficult to explain taking into account the expected pressure homogeneity in radio lobes favoured by large values of the sound speed. This argues against jet expansion as the mechanism to produce the elongated arc-shaped structure E. We suggest that it could be associated with a bow-shock triggered by changes in the injection power, in agreement with \cite{2003MNRAS.342..399G} and \cite{2022A&A...658A...5T}, or with instability patterns frozen in the disrupted structure, as suggested by our simulations (see Fig.~\ref{fig:sync}), or with a combination of both. Similar structures observed downstream \citep[structures D, C, B and A in the western jet;][]{2003MNRAS.342..399G} could also be interpreted in this way. Furthermore, the jet itself does not show evidence of expansion/recollimation before the appearance of the innermost ring \citep{2013ApJ...771...38O,2022A&A...658A...5T}, as expected if disruption would be caused by this process. 

Altogether, we understand that the jets keep collimation until the disruption point \citep[features W4 and E7 in the western and eastern jet, respectively,][]{2003MNRAS.342..399G} and therefore it cannot be caused by the arcs/rings and thus we can discard jet pinching as its trigger. Specific simulations using modulated injection and/or different instability modes should be run to try to reproduce those remarkable features; what our work shows is that the base set up to do this would necessarily imply a non-kinetically dominated flow.

\subsection{A hot jet}

Our interpretation of the nature of the jets in Hercules~A poses the question about the reason why the jets do not become completely kinetically dominated while keeping collimation at kiloparsec scales, as in the case of FRII jets, or show fast expansion due to mass-load and deceleration, as in the case of FRI jets, but remain hot, become disrupted, and generate bright lobes without hotspots.

The internal shocks produced by a relatively short-period variability in jet injection conditions \citep{2003MNRAS.342..399G,2022A&A...658A...5T} and the growth of the instability itself are effective agents in heating the jet plasma.
%The extended galactic density distribution contributes to keeping a larger lobe pressure that favours jet collimation, too. 
The large size of the galactic core contributes to keeping a large lobe pressure that favours jet collimation, too, preventing the adiabatic cooling of the flow.

The radius of the core of the density profile in this source is estimated to be $r_{\rm c}\,=\,120\,{\rm kpc}$, much larger than those of the galactic and group cores $1.2\,{\rm kpc}$ and $52\,{\rm kpc}$, respectively, that we have used in previous simulations of long-term jet evolution \citep{2014MNRAS.445.1462P,2019MNRAS.482.3718P,2022MNRAS.510.2084P}. Furthermore, the 
ambient pressure fall beyond $\sim 100 \,{\rm kpc}$ can also facilitate the fast lobe expansion in Hercules~A at these scales. Our simulations seem to confirm this, thus pointing towards an active role of the ambient medium, together with the intrinsic jet properties, in shaping the global morphology of the radio source.

It has also been shown that even a relatively mild toroidal field can favour jet collimation \citep[e.g.,][]{2022A&A...661A.117L}. By avoiding jet expansion, and therefore, the conversion of the stored internal energy into kinetic energy, the field
contributes to keep a relatively high internal energy density. A test to probe the relative relevance of each of the plausible main actors leading to the observed global morphology in Hercules~A (jet internal energy, magnetic field and ambient pressure profile) is left as future work.

\subsection{A HEG to LEG transition?}

Finally, it is interesting to relate the large-scale properties of the jet to the properties of the accretion disk in Hercules~A. In particular, the high values of internal energy that seem to explain their global morphology according to our results, thus linking the jet properties at formation with their final fate. \citet{2021A&A...647A..67B} investigated the relation between the accretion mode and the properties of the jet collimation region, by considering a sample of nearby sources classified as high- or low-excitation galaxies (HEGs or LEGs). The nuclei in the former class are thought to harbour standard accretion disks fueled by cold gas, whereas those in the latter are instead powered by radiatively inefficient, hot accretion flows \citep[][and references therein]{Heckman2014}. The authors suggested that jets in LEGs are mainly composed of a narrow relativistic hot spine anchored in the surroundings of the black hole ergosphere, while HEGs generate wider jets, likely due to the formation of an extended disk wind shielding the inner spine. Such winds may significantly contribute to the jet stability and result in the development of a kinetically dominated jet with an FRII morphology, as observed in the vast majority of HEGs. On the contrary, jets in LEGs would lack such mass-loaded, high-inertia, stabilizing shear layer, which prevents them from becoming kinetically-dominated and often results in the formation of an FRI morphology. Hercules~A presents a low-luminosity nucleus, and is optically classified as LEG \citep{Buttiglione2010}. In the scenario discussed above, we could thus speculate that this source forms a hot relativistic jet which, however, remains well collimated due to its peculiar ambient density/pressure profile, as explained before. 

While we have provided here a qualitative interpretation of the radio source properties in relation to the nucleus and the environment, it is necessary to note that a more complete description can be obtained by taking into account the complexity of the source evolution, in particular the evolution of nuclear properties. Our simulation of an internal energy dominated jet shows that the bow shock has propagated $\simeq 280\,{\rm kpc}$ in $\simeq 150\,{\rm Myr}$. %Therefore, and taking into account that the advance velocity becomes smaller with time, the injected jet would require several hundreds of millions of years to reach the observed size in Hercules~A. 
This long evolution time suggests that the jet could have initially propagated to a large distance while being more powerful and probably kinetically dominated \citep[see, e.g.][]{2019MNRAS.482.3718P,2022MNRAS.510.2084P}, and has later transitioned to a low-power regime due to a slow switching-off of the nucleus. 

Such a past, quasar-like activity is consistent with the analysis of the cavities and shock front observed in X-rays \citep{Nulsen2005}. Actually, these authors provide an age estimate for the activity burst of $60\,{\rm Myr}$ in which the size reached by the lobes is $\simeq 350\,{\rm kpc}$. This is a $25\%$ larger than the size reached by our simulation in a $40\%$ of the simulated time, revealing a clear dynamical contradiction. The authors obtain an estimated jet power of $10^{46}\,{\rm erg/s}$, i.e., a factor 2.5 larger than the one we have used in our simulation, which is consistent with the faster expansion given by their model. Therefore, either the model used by \citet{Nulsen2005} to estimate the source age fails to recover the detailed propagation dynamics, or indeed a quasar-like activity allowing for fast initial expansion is needed to reconcile both results. This mismatch can be explained by a slow transition in the accretion mode from efficient (HEG regime giving rise to the large scale structure) to inefficient (current LEG phase), probably caused by the depletion of the gas fueling galactic activity. The existence of this transition is in fact supported by the mismatch between the nuclear properties and the large scale structure of the source, as indicated by optical and radio data \citep{Buttiglione2010, Wu2020}.

\section{Conclusions} \label{sec:conc}
We present a numerical simulation that succeeds in reproducing the global morphology of Hercules~A-like jets, following an {\it a priori} hypothesis. The basic ideas behind the simulation are that 1) the spherical morphology of the lobes is pressure-driven, so the jet must be dominated by thermal (and perhaps magnetic) pressure, 2) the high lobe pressure keeps the jet collimated up to $\sim 100\,{\rm kpc}$, where it expands in the rapidly falling pressure/density WHIM, 3) the pressure in the lobe keeps high values due to the high internal energy flux through the jet and the opposition of the large ambient core, 4) the faster growth of KHI modes in hot flows favours jet disruption and the generation of the observed turbulent lobes, and 5) the observed arcs inside the lobes could be produced by non-linear amplitude helical instability modes, which are the responsible for jet disruption at $\sim 100-150\,{\rm kpc}$. 

We also speculate with the possibility that Hercules~A has transited from a HEG to a LEG, which could explain the large size of the radio galaxy, achieved by a powerful, probably kinetically dominated jet expected from HERGs, and the currently observed structure, which would imply a hot outflow like the one presented here. This interpretation would be in agreement with the implications of recent observational work on the properties of HERGs and LERGs, in terms of their size and the properties of the accretion flow from which they emerge \citep{2021A&A...647A..67B}.

Our hypothesis has proven valid to reproduce the large scale structure of the radio galaxy Hercules~A. Future work could be aimed to study the role of the magnetic field in this process, the possible transition phase and to confirm whether the generation of arcs can be fully explained by instability-induced shocks or whether periodicity in jet injection power is required.

\section*{Acknowledgements}
Computer simulations have been carried out in the Servei d'Inform\`atica de la Universitat de Val\`encia (Tirant). This work has been supported by the Spanish Ministry of Science through Grants PID2019-105510GB-C31/AEI/10.13039/501100011033, PID2019-107427GB-C33, and from the Generalitat Valenciana through grant PROMETEU/2019/071. JLM acknowledges support of a fellowship from ”La Caixa”
Foundation (ID 100010434). The fellowship code is LCF/BQ/DR19/11740030. The authors thank the anonymous referee for his/her constructive comments to the manuscript.

%%%%%%%%%%%%%%%%%%%%%%%%%%%%%%%%%%%%%%%%%%%%%%%%%%
\section*{Data Availability}

The data underlying this article will be shared on reasonable request to the corresponding author.



\bibliographystyle{mnras}
\bibliography{HerculesA} % if your bibtex file is called example.bib

%%%%%%%%%%%%%%%%%%%%%%%%%%%%%%%%%%%%%%%%%%%%%%%%%%

%%%%%%%%%%%%%%%%% APPENDICES %%%%%%%%%%%%%%%%%%%%%

%\appendix


%If you want to present additional material which would interrupt the flow of the main paper, it can be placed in an Appendix which appears after the list of references.

%%%%%%%%%%%%%%%%%%%%%%%%%%%%%%%%%%%%%%%%%%%%%%%%%%

% Don't change these lines
\bsp	% typesetting comment
\label{lastpage}
\end{document}